\documentclass[11pt]{article}
\usepackage{a4,latexsym,amsmath,amssymb,epsf}
\renewcommand{\slash}[1]{#1\!\!\!/}

\newcommand{\atfrac}[2]{\genfrac{}{}{0pt}{}{#1}{#2}}
\newcommand{\bea}{\begin{eqnarray}}
\newcommand{\beas}{\begin{eqnarray*}}
\newcommand{\ena}{\end{eqnarray}}
\newcommand{\enas}{\end{eqnarray*}}
\newcommand{\dsdreigluonfig}{1}
\newcommand{\tmatrixfig}{2}
\newcommand{\softfig}{3}
\newcommand{\dsbnullgluonfig}{4}
\newcommand{\dsdreighostfig}{5}
\newcommand{\dszweifermfig}{6}
\newcommand{\dsdreifermfig}{7}
\allowdisplaybreaks[2]

\begin{document}
\pagestyle{empty}
\renewcommand{\theequation}{\arabic{section}.\arabic{equation}}
\begin{flushright}
MS--TPI--97--4\\
December 1997\\
hep-th/9808152
\mbox{}\\
\mbox{}\\
\mbox{}\\
\end{flushright}
\begin{center}
{\Large \bf{Extended Iterative Scheme for QCD:}}\\
{\Large \bf{Three-Point Vertices}}
\end{center}
\mbox{}\\
\mbox{}\\
\mbox{}\\
\begin{center}
L.\ Driesen, J.\ Fromm, J.\ Kuhrs, and M.\ Stingl\\
Institute for Theoretical Physics I, University of M\"unster\\
D-48149 M\"unster (Westf.), Germany
\end{center}
\mbox{}\\
\mbox{}\\
\mbox{}\\
\mbox{}\\
\mbox{}\\
{\large \bf{Abstract:}} \
In the framework of a generalized iterative scheme introduced previously 
to account for the non-analytic coupling dependence associated with the 
renormalization-group invariant mass scale $\Lambda$, we establish the 
self-consistency equations of the extended Feynman rules ($\Lambda$-modified 
vertices of zeroth perturbative order) for the three-gluon vertex, the two
ghost vertices, and the two vertices of massless quarks. Calculations 
are performed to one-loop-order, in Landau gauge, and at the lowest 
approximation level $(r=1)$ of interest for QCD. We discuss the phenomenon 
of compensating poles inherent in these equations, by which the formalism 
automatically cancels unphysical poles on internal lines, and the
role of composite-operator information in the form of equation-of-motion
condensate conditions. The observed near decoupling of 
the four-gluon conditions permits a solution to the 2-and-3-point conditions
within an effective one-parameter freedom. There exists a parameter range 
in which one solution has all vertex coefficients real, as required for 
a physical solution, and a narrower range in which the transverse-gluon 
and massless-quark propagators both exhibit complex-conjugate pole pairs.
\newpage\mbox{ }   \pagestyle{empty}   \newpage
\pagestyle{plain}
\setcounter{page}{1}

\section{Summary of the extended iterative scheme}
\setcounter{equation}{0}

The spontaneous emergence of a renormalization-group (RG) invariant mass 
scale $\Lambda$ \cite{1} from the renormalization process is arguably the 
most important nonperturbative effect in strictly renormalizable quantum 
field theories, since this quantity sets the scale for all dimensionful 
observables (except those dominated by heavy extraneous masses). Its 
coupling dependence,
\begin{equation} \label{lambda}
       (\Lambda^2)_R =\nu^2\exp\Bigg\{-2\int\limits^{g(\nu)}
                 \frac{dg'}{\left[\beta(g')\right]_R}\Bigg\}
              =\nu^2\exp\left\{-\frac{(4\pi)^2}{\beta_0 g(\nu)^2}
             \left[1+{\cal O}(g^2)\right]_R\right\}, 
\end{equation}
(where $R$ denotes a renormalization scheme and $\nu$ the arbitrary 
renormalization scale within $R$, and where $\beta_0>0$ in an asymptotically 
free theory) is non-analytic in a way that will always remain invisible in 
a perturbation expansion around $g^2=0$. Moreover, the several known 
obstructions \cite{2} to the existence and uniqueness of a Borel transform 
of the perturbation series all have to do, at least qualitatively, with the 
presence of this scale. There exists, therefore, the intriguing {\it possibility that accounting systematically for the $\Lambda$ dependence of correlation 
functions may be the minimal step beyond perturbation theory needed to 
define a strictly renormalizable theory uniquely}. (Here "$\Lambda$ 
dependence" does not, of course, refer to a mere reparametrization of the 
perturbation series, as obtained by solving (\ref{lambda}) for $g$ in terms 
of $\Lambda$, or equivalently by leading-logarithms resummation. Genuinely 
nonperturbative $\Lambda$ dependence, exemplified by the way vacuum 
condensates occur in operator-product expansions (OPE), is typically 
polynomial or inverse-polynomial).

The present paper elaborates on a specific scheme, outlined earlier in this 
journal \cite{3}, of accounting systematically for the $\Lambda$ dependence, 
under the restriction (which in an asymptotically free theory turns out to 
be a weak one) that the known standard technique of renormalization remain 
applicable with at most inessential modifications. This scheme takes the 
form of an {\it extended iterative solution} to the integral equations for 
correlation or vertex functions, starting from a set of extended Feynman 
rules for the superficially divergent basic vertices of the theory, as 
distinct from the ordinary Feynman rules (bare vertices) $\Gamma^{(0)pert}$, 
whose iteration generates the perturbative series. These extended vertices 
are quantities of zeroth order $(p=0)$ with respect to the perturbative 
$g^2$ dependence but contain a ``seed dependence'' on the nonanalytic scale 
(\ref{lambda}), approximated systematically. Under the combined requirements 
of globality (in order to be applicable in loop integrals, the approximation 
must in principle be valid over the entire momentum range) and of the 
preservation of power counting (as a basic prerequisite of standard 
renormalization technique), the choice of approximating functions is 
remarkably unique: they must be functions {\it rational} with respect to 
$\Lambda$ and therefore (since $\Lambda$ is dimensionful) also with 
respect to momentum variables. For the representation of the $(p=0)$ 
nonperturbative $\Lambda$ dependence, rational approximants perform the 
same basic role that polynomial approximants play for the perturbative 
$g^2$ dependence. The poles of these rational functions, together with the 
numerator zeroes, will in general form discrete approximations to branch 
cuts of the vertices in their complex-momentum planes. We follow \cite{3} in 
denoting the extended rules by $\Gamma^{[r,0]}$, with $r$ the (denominator) 
degree of rational approximation. At each level $r$, the $\Gamma^{[r,0]}$
approach the $\Gamma^{(0)pert}$ in the ``perturbative limit'' $\Lambda
\rightarrow 0$; on the other hand their sequence as $r$ increases may
be viewed as an analytic continuation-through-resummation of the
zeroth-order terms of the OPE.

The central problem of such a method is to demonstrate dynamical 
self-consistency of these extended Feynman rules. The integral equations 
(Dyson-Schwinger, or DS, equations) for the set $\Gamma$ of proper vertex 
functions $\Gamma_N$, where $N$ stands summarily for the number and types of 
external legs, take the schematic form
\begin{equation} \label{functional-dse}
            \Gamma_N=\Gamma_N^{(0)pert}+\left(\frac{g_0}{4\pi}\right)^2
                   \Phi_N[\Gamma^{(0)pert},\Gamma],
\end{equation}
where $g_0$ is the bare gauge coupling, and $\Phi_N$ a set of nonlinear 
dressing functionals, containing loop integrals over combinations of 
$\Gamma$'s. With the inhomogeneous terms always given by the bare 
$\Gamma^{(0)pert}$, iterating instead around a nonperturbatively modified 
set of starting functions $\Gamma^{[r,0]}$ can work only if the interaction 
terms $\Phi_N$, at each order $l$ of the iteration, not only generate an 
$l$-th order, "quasi-perturbative" power correction in $g^2(\nu)$, but also 
reproduce the nonperturbative parts, $\Gamma^{[r,0]}-\Gamma^{(0)pert}$, of 
the new zeroth-order input: as compared to perturbation theory, the solution 
must be able to "establish its own zeroth order". To have the functionals 
$\Phi_N$, in spite of their $g_0^2$ prefactor, produce certain terms of 
{\it zeroth} order in $g^2$ is not trivial. Moreover, for the method to 
be practical, the number of extended Feynman rules should remain finite 
(and small), as in perturbation theory, and in view of the notorious 
infinite hierarchical coupling in eqs.\ (\ref{functional-dse}) -- with 
each $\Phi_N$ coupling to $\Gamma$'s up to $\Gamma_{N+1}$ or even 
$\Gamma_{N+2}$ --, this is again nontrivial.

The mechanism discussed in \cite{3} for simultaneously ensuring both 
objectives exploits the structure (\ref{lambda}) of the scale $\Lambda$ in 
conjunction with the renormalizable divergence structure of the theory. To 
briefly describe its main line, let
\begin{equation} \label{num-coef}
             \left\{c\right\}^{[r]}=\left\{c_{r,1},c_{r,2},
                   \ldots c_{r,k_r}\right\}
\end{equation}
be the complete set of $k_r$ numerator coefficients of the level-$r$ 
rational approximants $\Gamma_N^{[r,0]}$ for all superficially divergent 
vertices (of which, we recall, there are seven in covariantly quantized QCD), 
and let
\begin{equation} \label{den-coef}
              \left\{d\right\}^{[r]}=\left\{d_{r,1},d_{r,2},
                   \ldots d_{r,m_r}\right\}
\end{equation}
be the complete set of $m_r$ denominator zeroes (pole positions) in units 
of $\Lambda^2$. Both $c$'s and $d$'s are dimensionless, real numbers. Then
\begin{equation} \label{general-ansatz}
              \Gamma_N^{[r,0]}=\Gamma_N^{(0)pert}+\Delta_N^{[r]}
                \big(\{c\}_N^{[r]},\{d\}_N^{[r]};\Lambda\big),
\end{equation}
in a notation suppressing momenta and all other variables not immediately 
pertinent to the argument. Here $\{c\}_N^{[r]}$ denotes the subset of 
$\{c\}^{[r]}$ appearing in the vertex $\Gamma_N$, etc. Upon evaluating, say, 
the first iteration (one-loop order, $l=1$) of eq. (\ref{functional-dse}), 
dimensionally regularized in $D=4-2\epsilon$, with the functions 
(\ref{general-ansatz}) as input, one obtains after some algebraic 
decomposition,
\begin{align}
            \left[\frac{g_0(\epsilon)\nu_0^\epsilon}{4\pi}\right]^2
             \Phi_{N,\epsilon}^{(l=1)}[\Gamma^{(0)pert},\Gamma^{[r,0]}] 
    \, = \, \Pi(\epsilon) & \cdot \Delta_N^{[r]}\big(\{C(\{c\}^{[r]},
         \{d\}^{[r]})\}^{[r]},\{d\}_{N'\neq N}^{[r]};\Lambda\big) \nonumber\\
         + \left[\frac{g_0(\epsilon)\nu_0^\epsilon}{4\pi}\right]^2
       & \cdot \left\{\Xi_N^{(1)}\big(\{c\}^{[r]}\big)\frac{1}{\epsilon}
           +\Gamma_N^{[r,1)}\big(\{c\}^{[r]},\{d\}^{[r]};\Lambda\big)
                      +{\cal O(\epsilon)}\right\}. \label{functional-iter}
\end{align}
Here $\{C\}^{[r]}$ is a set of nonlinear algebraic expressions in the 
input coefficients (\ref{num-coef}/\ref{den-coef}), while 
$\{d\}_{N'\neq N}^{[r]}$ denotes the subset of denominator roots 
(\ref{den-coef}) in the vertices $\Gamma_{N'}$ other than $\Gamma_N$ to 
which $\Phi_N$ provides coupling. The appearance of the first term on the 
r.h.s.\ of (\ref{functional-iter}) is nontrivial: it occurs only if all 
seven basic vertices are treated by mutually consistent, nonperturbative 
approximants of the same level $r$. This term appears with a prefactor,
\begin{equation} \label{pi}
           \Pi=\left[\frac{g_0(\epsilon)}{4\pi}\right]^2
               \frac{1}{\epsilon}\left(\frac{\Lambda_\epsilon^2}
               {\nu_0^2}\right)^{-\epsilon},
\end{equation}
which, by virtue of an exact RG identity, is independent of the renormalized 
coupling $g(\nu)$, and finite as $\epsilon\to 0$:
\begin{equation} \label{pi-eps}
           \Pi(\epsilon)=\frac{1}{\beta_0}\left[1+{\cal O}
              (\varepsilon,\varepsilon\ln\varepsilon)\right],
                  \quad\mathrm{independent}\;\mathrm{of}\;g^2.
\end{equation}
Here $\beta_0$ is the leading beta-function coefficient of 
eq.\ (\ref{lambda}). Note how in this exact result one coupling factor 
$(g_0^2)$ and one divergence factor $\left(\frac{1}{\epsilon}\right)$ get 
"eaten" to produce a coupling-independent and finite quantity.

It therefore becomes possible to reproduce analytically the nonperturbative 
part, $\Delta_N^{[r]}$, of the zeroth-order input (\ref{general-ansatz}) by 
imposing the matching or self-consistency conditions,
\begin{equation} \label{den-sc}
             \{d\}_{N'\neq N}^{[r]}=\{d\}_N^{[r]}\quad(\mathrm{all}\;N),
\end{equation}
which says that all basic vertices must exhibit one common set of 
denominator zeroes (still differing, however, for different types of 
external legs, or basic fields), and
\begin{equation} \label{num-sc}
             \frac{1}{\beta_0}C_i^{[r]}\big(\{c\}^{[r]},\{d\}^{[r]}\big)
                =c_{r,i}\quad(i=1,\ldots k_r),
\end{equation}
which ensures reproduction of numerator structures.

It is crucial that the nonperturbative terms establish themselves in a 
{\it finite} manner, since in this way one avoids the introduction of 
nonlocal counterterms, and thus preserves another basic element of standard 
renormalization. It is equally crucial that the above mechanism, as shown by 
the last two factors of (\ref{pi}), is {\it tied to the loop divergences} of 
the integral equations: this gives the superficially divergent vertices a 
privileged position, such that formation of nonperturbative $\Delta_N$'s
remains rigorously restricted to these vertices. In spite of the infinite 
hierarchical coupling, {\it the number of extended Feynman rules does not 
proliferate}, and in fact remains the same as for the bare vertices.

In what follows, we focus exclusively on this self-consistency process for 
the generalized Feynman rules, and therefore refer the reader to \cite{3} 
for what needs to be said about the last term of eq.\ (\ref{functional-iter}) 
-- representing the $p=1$ quasi-perturbative correction -- and about the 
perturbative ``boundary condition'' and 
essentially standard renormalization procedure it requires. We only note, 
for later use, that the condition of having the remaining divergence exactly 
equal to the perturbative one would require $\Xi_N^{(1)}$ to be of the form
\begin{equation} \label{pert-sc}
               z_N^{(1)}\Gamma_N^{(0)pert},
\end{equation}
with $z_N^{(1)}$ the one-loop coefficient of $(g/4\pi)^2/\epsilon$ in the 
perturbative renormalization constant $Z_N$ for the Vertex $\Gamma_N$ -- a 
condition which may impose extra constraints on $\{c\}^{[r]}$ that for low 
$r$ may be satisfiable only within approximation errors.

The exploration of this extended-iterative scheme represents a calculational 
program of some length. It was begun in \cite{3} with an illustrative 
derivation of eq.\ (\ref{functional-iter}) and of matching conditions 
(\ref{num-sc}) at $r=1$ and $l=1$ for the transverse two-gluon vertex. While 
calculations of the ghost and fermion two-point functions follow essentially 
the same pattern, those for the higher superficially divergent vertices, 
with $N=3$ and $4$, are not straightforward extensions to more kinematical 
variables. As a result of the subtle interplay between the coupled DS 
equations, they reveal a whole array of new aspects and intricacies. We 
therefore plan to present and discuss the $r=1$, $l=1$ self-consistency 
calculations for these vertices in several parts. In the present paper, we 
focus on the remaining superficially divergent vertices of the gauge, ghost, 
and massless-quark sectors up to $N=3$: the three-gluon vertex $\Gamma_{3V}$, 
ghost vertices $\Gamma_{G\bar G}$ and $\Gamma_{GV\bar G}$, where $V$ and $G$ 
label vector (=gluon) and ghost external legs, and fermion vertices 
$\Gamma_{F\bar F}$ and $\Gamma_{FV\bar F}$. In the companion paper \cite{4} 
we will deal with the {\it four-gluon vertex} $\Gamma_{4V}$, the highest 
superficially divergent vertex, which is particularly complicated both 
kinematically and in its DS equation. In these two papers, consideration 
of the fermion (quark) functions will be restricted -- as were the $N=2V$ 
calculations of \cite{3} -- to the case of massless quarks, where the 
fermionic mass scales, too, are simply multiples of $\Lambda$. For massive
fermions, the presence of "extraneous" RG-invariant mass scales not having
the structure (\ref{lambda}) causes additional complications with which
we plan to deal separately.

In section 2 of the present article, we recall the DS equation 
for the $\Gamma_{3V}$ vertex. While 
summarizing known material, this section seems necessary to establish 
notation and a precise starting point for the subsequent discussion. In the 
present program we deal exclusively with the "ordinary" DS equations, 
without additional Bethe-Salpeter resummations in their interaction terms, 
in which the distinguished leftmost external line always runs into a 
{\it bare} vertex. Section 2.2 focuses on the phenomenon of compensating
poles in the $3V$ equation where they make their first appearance. While at 
first sight a merely technical point, these turn out to be an important 
structural element, by which the formalism automatically prevents the 
appearance of "wrong" poles on internal lines. A systematic account of these 
leads to the rearranged integral equation of section 2.3, whose terms now 
exhibit an {\it extended-irreducibility property}. It is the rearranged 
equations that form the most convenient framework for the self-consistency 
problem of the extended Feynman-rules. Extraction of the matching 
conditions (\ref{den-sc}-\ref{pert-sc}) at the lowest level of rational 
approximation of interest for QCD ($r=1$) and at one loop ($l=1$) is 
discussed in sect.\ 2.4 for the 3-gluon vertex. Section 3 considers the 
equations for the two ghost vertices, and section 4 the equations for the 
two fermion vertices in the massless case. 

The combined 2-plus-3-point
self-consistency system is discussed in sect.\ 5\ . A noteworthy result,
which will continue to hold after inclusion of 4-gluon-conditions, is
that the set of denominator coefficients (\ref{den-coef}), while restricted by
(\ref{den-sc}), is not fully determined by the divergent parts of DS loops,
and that composite-operator information, in the form of equation-of-motion
condensate conditions ( DS equations at coincident spacetime points )
is required at this point to complement the usual equations.
On the other hand, the system is found to nearly decouple from the 
4-gluon one, so that a solution without 4-gluon equations is possible
with an effective one-parameter freedom. The results, when compared 
to the more restricted and heuristic attempt of refs. \cite{5} 
for a pure-gluon theory, will be seen to represent significant progress, 
and in particular to include the existence of a parameter range where 
one of the several solutions to the nonlinear system has all zeroth-order
vertices entirely real in the Euclidean, as required for a physical 
solution.

As in \cite{3}, all calculations are performed {\it for the Euclidean 
theory, and in Landau gauge}. The Landau gauge provides some welcome 
reduction of the considerable complexity of loop computation with the 
extended Feynman rules: here the two ghost vertices turn out to remain 
perturbative, and calculations can be restricted to amplitudes with only 
transverse (if any) gluon legs, which then form a closed DS problem. This 
has the obvious disadvantage that nothing can be inferred as yet about the 
approximate saturation or violation of Slavnov-Taylor identities, which we 
recall are statements about amplitudes with at least one {\it longitudinal} 
gluon leg, and which in the low orders of an iterative scheme are not, of 
course, expected to be exactly self-consistent. However, since the physical 
degrees of freedom of the gauge field are in the transverse sector, one may 
expect the essential parts of the nonperturbative structure to develop here 
(as is obviously true for the 2-gluon function). Also, we do not yet use the 
most general color-and-Lorentz-tensor structure of the $4V$ vertex, which 
would lead to calculations of prohibitive length, but restrict our study to 
a theoretically motivated tensor subset capable of dynamical 
self-consistency. Nor do we consider the quasi-perturbative corrections, 
$\left(\frac{g}{4\pi}\right)^2\Gamma_N^{[1,1)}$ in the notation of eq.\ 
(\ref{functional-iter}). While all these questions are interesting in 
themselves, they must form subjects of future study.

To the extent that the present scheme uses the DS equations as a framework, 
its purpose is not to provide exact numerical solutions of DS equations at 
low levels of decoupling; for a review of the work in this direction the 
reader is referred to \cite{6}. Here the aim is to develop an analytic 
approximation method that provides some insight into the nonperturbative 
{\it coupling} structure, and in particular to identify the precise 
mechanism, connected with the divergence structure, by which the scale 
(\ref{lambda}) establishes itself in correlation functions. In particular, 
such a scheme allows qualitative changes in the elementary propagators -- 
the appearance of zeroth-order, finite, real or complex mass shifts -- 
to be followed in a more transparent fashion.

\section{The three-gluon vertex equation}
\subsection{Integral equation and input}
\setcounter{equation}{0}

The DS equation for the proper three-vector vertex $g_0\Gamma_{3V}$ in 
Euclidean momentum space is written diagrammatically in 
Fig.\ \dsdreigluonfig. The form shown is a compact but hybrid one: most 
terms on the r.h.s.\ have not been resolved down to the level of proper 
vertices, but feature connected and amputated functions $T'$ which are 
one-particle irreducible (1PI) only in the horizontal channel of the 
diagram, while otherwise still containing reducible (1PR) terms. Thus in 
term $(A)_3$ of Fig.\ \dsdreigluonfig, the four-gluon $T$ matrix $T'_{4V}$
for the horizontal channel ( denoted there as $T'_{s}$ ) is to be 
decomposed further as in Fig.\ \tmatrixfig:
\begin{align}
      T_{4V}' & = T_{4V}-A_1 \label{t4-prime} \\ 
      T_{4V}  & = A_1+A_2+A_3+\Gamma_{4V}. \label{t4}
\end{align}
Here $\Gamma_{4V}$ is the proper, fully 1PI, four-gluon vertex, 
while $A_1$ and $A_2$, $A_3$ are dressed one-gluon reducible terms 
in the horizontal channel and the two crossed channels, 
respectively. Analogous relations apply to the ghost-antighost-gluon-gluon 
and quark-antiquark-gluon-gluon $T'$ matrices, $T_{G\bar GVV}'$ 
and $T_{F\bar FVV}'$, of terms $(B)_3$ and $(E)_3$ respectively.

The "standard" form of the $\Gamma_{3V}$ equation in Fig.\ \dsdreigluonfig\  
displays the characteristic asymmetry, common to all DS equations, of having 
the leftmost external leg always ending in a bare vertex, while the other 
legs run into dressed vertices. This structure is at the core of a problem 
plaguing all treatments (and not just the present approximation method) of 
vertices with $N\ge 3$: while the exact solution of the equation may be 
known to have a certain Bose or Fermi symmetry, the equation does not 
display this symmetry manifestly, and approximate solutions to it therefore 
usually fail to exhibit the full desired symmetry. Enforcing the symmetry by 
imposing extra conditions on the vertex coefficients leads to 
{\it overdetermination} in the self-consistency equations. In the framework 
of an iterative solution, "trivial" symmetrizations could of course be used 
to cure this problem, but at the expense of depriving oneself of an important 
{\it indicator of the overall error} at the level of approximation considered.

Since we will be working throughout at the one-loop ($l=1$) level, 
characterized by a single $D$-dimensional momentum integration in the 
dressing functional $\Phi_{3V}$, the term $(D)_3$ of Fig.\ \dsdreigluonfig\  
with two DS loops does not yet contribute (its contribution to the l.h.s.\ 
of eq.\ (\ref{num-sc}) will be of order $(1/\beta_0)^2$).

The {\it input} for the self-consistency calculation must consist of the 
Euclidean extended Feynman rules $\Gamma_N^{[r,0]}$, at the same level $r$ 
of rational approximation, for all seven superficially divergent vertices. 
With the exception of $\Gamma_{4V}^{[r,0]}$, whose $r=1$ form will be 
detailed in \cite{4}, these have been listed in \cite{3}. Here we need to 
recall only two elements carrying special restrictions. First, the 
gluon-propagator rule, $D^{[r,0]}=-\big(\Gamma_{2V}^{[r,0]}\big)^{-1}$, will 
be simplified throughout by adopting the Landau ($\xi=0$) gauge fixing. Then
\begin{align}
        \big(D^{[r,0]}(k)\big)^{\mu\nu}_{ab} & = 
        \delta_{ab}t^{\mu\nu}(k)D_T^{[r,0]}(k^2), \label{gl-prop-color} \\
         t^{\mu\nu}(k) & = 
            \delta^{\mu\nu}-\frac{k^\mu k^\nu}{k^2}, \label{tr-proj}
\end{align}
where, at the $r=1$ level,
\begin{align}
          D_T^{[1,0]}(k^2) & = \Big[ k^2+u_{1,1}\Lambda^2
              +\frac{u_{1,3}\Lambda^4}{k^2+u_{1,2}
                       \Lambda^2}\Big]^{-1} \label{gl-prop} \\
       & = \frac{k^2+u_{1,2}\Lambda^2}{(k^2+\sigma_{1,1}\Lambda^2)
             (k^2+\sigma_{1,3}\Lambda^2)}. \label{gl-prop-sigma}
\end{align}

Second, the general {\it color} structure of the $\Gamma_{3V}$ vertex itself, 
whose self-reproduction we examine in this section,
\begin{equation}
       (\Gamma_{3V})_{abc}=if_{abc}\Gamma_{(f)}+d_{abc}\Gamma_{(d)},
\end{equation}
will be simplified from the outset to a pure $f_{abc}$ structure, i.e.\ 
one puts
\begin{equation}
               \Gamma_{(d)} \approx 0
\end{equation}
and omits the $f$ on $\Gamma_{(f)}$. The reason is that in the much more 
complicated color structure of $\Gamma_{4V}$ discussed in \cite{4}, we will 
disregard those color-basis tensors that would feed the $d_{abc}$ portion 
through the $(A)_3$ term of Fig.\ \dsdreigluonfig. It is conceivable that 
some $d_{abc}$ structure could be made to self-reproduce through the other 
one-loop terms of Fig.\ \dsdreigluonfig\  alone, but since we view the seven 
basic DS equations as an interrelated whole, it does not seem consistent to 
us to keep one source of such terms and neglect the other. 
The tensor structure then is
\begin{align}
           \big(\Gamma_{3V}(p_1,p_2,p_3)\big)^{\rho\kappa\sigma}_{abc}
                    =if_{abc}\Big\{  
                  \quad  & \delta^{\kappa\sigma}(p_2-p_3)^\rho 
                                F_0(p_2^2,p_3^2;p_1^2) \nonumber\\
              + \, & \delta^{\sigma\rho}(p_3-p_1)^\kappa 
                                F_0(p_3^2,p_1^2;p_2^2) \nonumber\\
              + \, & \delta^{\rho\kappa}(p_1-p_2)^\sigma 
                                F_0(p_1^2,p_2^2;p_3^2) \nonumber\\
              + \, & (p_2-p_3)^\rho(p_3-p_1)^\kappa(p_1-p_2)^\sigma 
                              F_1(p_1^2,p_2^2,p_3^2) \nonumber\\
              + \, \big[ \textrm{ 10 terms not } & \textrm{contributing }
                     \textrm{to totally transverse vertex } \big]
              \Big\}.                                    \label{gl-3-vert}
\end{align}
At level $r=1$ (and only at $r=1$), the invariant functions $F_0^{[1,0]}$ 
and $F_1^{[1,0]}$ are conveniently written in a form fully decomposed into 
partial fractions,
\begin{align}
          F_0^{[1,0]}(p_1^2,p_2^2;p_3^2) =
                 & 1+x_{1,1}\big(\Pi_1+\Pi_2\big)+\Big(x_{1,2}+
          \frac{x_{1,2}'}{\Pi_3}\Big)\Pi_1\Pi_2+x_{1,3}\Pi_3 \nonumber\\
                 & +\Big[\Big(x_{1,4}+\frac{x_{1,4}'}{\Pi_2}\Big)\Pi_1
       +\Big(x_{1,4}+\frac{x_{1,4}'}{\Pi_1}\Big)\Pi_2\Big]
                    \Pi_3+x_{1,5}\Pi_1\Pi_2\Pi_3,  \label{f0}
\end{align}
\begin{align} \label{f1}
          F_1^{[1,0]}(p_1^2,p_2^2,p_3^2) = \frac{1}{\Lambda^2}\Big[
                x_{1,6}\big(\Pi_1\Pi_2+\Pi_2\Pi_3+\Pi_3\Pi_1\big)
                          +x_{1,7}\Pi_1\Pi_2\Pi_3\Big]
\end{align}
featuring the building blocks
\begin{equation} 
           \Pi_i=\frac{\Lambda^2}{p_i^2+u_{1,2}'\Lambda^2}
                            \quad (i=1,2,3). \label{poles}
\end{equation}
Here we anticipate that the DS self-consistency conditions (\ref{den-sc}) 
will enforce $u_{1,2}'=u_{1,2}$, i.e., a common denominator factor in all 
gluonic vertices. The {\it factorized} (with respect to the 3 variables 
$p_i^2$) denominator structure of these approximants may be viewed as 
arising from a triple-spectral representation,
\begin{equation} \label{spectral}
            F_{0,1}=\frac{1}{\pi^3}\int dz_1dz_2dz_3\frac{\rho_{0,1}
       (z_1,z_2,z_3,p_1^2,p_2^2,p_3^2)}{(z_1-p_1^2)(z_2-p_2^2)(z_3-p_3^2)},
\end{equation}
(where the spectral functions are still allowed some {\it polynomial} 
$p_i^2$ dependence) by a discrete approximation. At $r=1$, it exhibits a 
single denominator zero, $p_i^2=-u_{1,2}\Lambda^2$, in all three variables.

Although the approximation of branch-cut structures by poles is an old 
technique, a casual look might still suggest a danger here that these poles 
in vertices could somehow take on a life of their own and roam the formalism 
as unphysical particles. This does not happen, for two reasons. First, DS 
self-consistency will transfer these poles down to the two-point vertices 
(negative-inverse propagators), so that the propagators themselves will 
develop {\it zeroes} at these pole positions, as seen in 
(\ref{gl-prop-sigma}). The unamputated, connected Green functions, which are 
the quantities having physical interpretation as propagation amplitudes, 
will therefore be nonsingular at these positions in the squared momenta of 
their {\it external} lines, and no S-matrix elements for unphysical 
particles of masses $u_{1,2}\Lambda^2$ will arise, as emphasized already 
in \cite{3}. Second, and perhaps more remarkably, the formalism will also 
automatically cancel the poles of type (\ref{poles}) when they arise on 
certain {\it internal} lines, so that there will be no Cutkosky 
discontinuities corresponding to production of such objects. To demonstrate 
this in the simplest context will be the subject of the next subsection.

\subsection{"Compensating" poles in $\Gamma_{4V}$}

The phenomenon of automatic cancellation of superfluous poles on internal 
lines seems to have been noted first by Jackiw and Johnson and by Cornwall 
and Norton \cite{7} in 1973. While there are obvious technical differences 
between their (Abelian) models and the present QCD study ( in particular,
their poles arise in longitudinal-vector channels, whereas here they
appear in the transverse-gluon sector ), this mechanism is 
very much the same in both cases.  

Consider the $\Gamma_{3V}$ equation of Fig.\ \dsdreigluonfig\  in order 
$[r,0]$, and compare residues of both sides at the poles in the variable 
$p_2^2$ of the leftmost leg. In a partial-fraction decomposition with 
respect to $p_2^2$, the l.h.s.\ can be written
\begin{align} 
       \Gamma_{3V}^{[r,0]} \
                 & =\  \sum_{n=1}^r B_n^{[r]}(p_3,p_1)\frac{\Lambda^2}
                 {p_2^2+u_{r,2n}\Lambda^2}\ +\ B_0^{[r]}(p_3,p_1) \nonumber\\
         & +\  [ \mbox{ terms with } (p_2^2)^1,(p_2^2)^2,
                 \ldots (p_2^2)^r\,]. \label{gamma-3v-partial}
\end{align}
The compact notation suppresses all tensor structure. All terms displayed 
still have rational structure in $p_3^2$, $p_1^2$. The terms in the second 
line are allowed \cite{3} as long as only the conditions of overall 
asymptotic freedom and preservation of perturbative power counting are 
imposed, but will in fact turn out to be more strongly restricted. The 
residue at $p_2^2=-u_{r,2n}\Lambda^2$ for the l.h.s.\ is then $\Lambda^2 
B_n$ ($n=1,\ldots r$).

For the r.h.s., to keep the argument as simple as possible, we invoke for 
the moment all simplifications available for our specific calculation 
according to subsect.\ 2.1, i.e.\ we disregard terms $(D)_3$, $(E)_3$, and 
$(B)_3$ (the general case will be outlined in subsect. 2.3). Then the only 
denominator structure with respect to $p_2^2$ must come from the 
${T'}_{4V}^{[r,0]}$ amplitude of term $(A)_3$, for which 
$p_2^2=(p_3+p_1)^2=s_E$ is the (Euclidean) Mandelstam variable in the 
horizontal channel. According to general structure theorems on correlation 
functions \cite{8}, residues at poles in such a variable must factorize with 
respect to the two sides of the channel, or be at most sums of factorizing 
terms. Thus if $q_1$, $q_3$ are the momenta of the loop gluons of term 
$(A)_3$, the $T'_{4V}$ amplitude in the vicinity of 
$s_E=-u_{r,2n}\Lambda^2$ must behave as
\begin{equation} \label{t-comp-pole}
         {T'}_{4V}^{[r,0]}=\frac{\Psi_n^T(q_1,q_3)\Psi_n(p_3,p_1)}
                  {s_E+u_{r,2n}\Lambda^2}+[\mbox{ regular terms }].
\end{equation}
The partial Bose symmetry of $T'_{4V}$ implies that both residue factors 
must be given by the same function or column vector of functions, $\Psi_n$. 
Note that (\ref{t-comp-pole}) in no way contradicts the 1PI property of 
$T'_{4V}$ in the horizontal channel: the pole factor is not a propagator 
of any of the {\it elementary} fields of the theory, so the term is
technically allowed to appear ( as would, e.g., a bound-state pole ) not only
in $T'_{4V}$ but in fact in the fully 1PI piece, $\Gamma_{4V}$, of eq.\ 
(\ref{t4}). However, the observation does suggest a natural enlargement of 
the notion of reducibility, as will be discussed below.

The residue comparison shows that $\Psi_n$ must be proportional to $B_n$,
\begin{equation} \label{psi}
      \Psi_n(p_3,p_1)=M_nB_n(p_3,p_1),
\end{equation}
with $M_n$ some matrix, and that (again in compact notation)
\begin{equation} \label{vert-pole-sc}
            \Big\{g_0^2\frac{1}{2}\int\Gamma_{3V}^{(0)pert}
      DDB_n\Big\}_{p_2^2=-u_{r,2n}\Lambda^2}\cdot M_n^TM_nB_n=\Lambda^2B_n.
\end{equation}
But here the brackets on the l.h.s.\ are already fixed from a lower stage of 
the DS problem: the self-reproduction conditions for the poles of the gluon 
self-energy $\Delta_{2V}(p_2)$. Those enforce, under the simplifications 
adopted,
\begin{equation} \label{prop-pole-sc}
         \Big\{g_0^2\frac{1}{2}\int\Gamma_{3V}^{(0)pert}
        DDB_n\Big\}_{p_2^2=-u_{r,2n}\Lambda^2}=-u_{r,2n+1}\Lambda^2t(p_2),
\end{equation}
$t$ being the transverse projector of eq.\ (\ref{tr-proj}), and 
$u_{r,2n+1}$ the dimensionless residue parameter in $\Gamma_{2V}^{[r,0]}$ 
exemplified, for $r=1$, by the $u_{1,3}$ of eq.\ (\ref{gl-prop}). We 
conclude that, first,
\begin{equation} \label{bn-trans}
       B_n^{[r]}(p_3,p_1)=t(p_3+p_1)B_n^{[r]}(p_3,p_1)\quad (n=1,\ldots r),
\end{equation}
an information more detailed than that of the all-transverse projection 
(\ref{gl-3-vert}): nonperturbative denominator structure develops only in 
the variables of {\it transverse} gluon legs. Second,
\begin{equation} \label{m-def}
       M_n^TM_n=-\frac{1}{u_{r,2n+1}}t(p_2).
\end{equation}
What has been learned is that {\it as a consequence of the lower} 
(two-point and three-point) {\it equations alone}, ${T'}_{4V}^{[r,0]}$ and 
$\Gamma_{4V}^{[r,0]}$ must contain a term of the form
\begin{equation}                                      \label{2v2v-comp-pole}
       - \ \big(C_1^{[r]}\big)_{2V,2V}=\,-\,\sum_{n=1}^rB_n^{[r]} \,
       \frac{(u_{r,2n+1})^{-1}t(P)}{P^2+u_{r,2n}\Lambda^2} \, B_n^{[r]}
\end{equation}
in the total momentum $P$, with $P^2=s_E$, of its horizontal channel. But 
$\Gamma_{4V}$ is fully Bose symmetric, and therefore must contain analogous 
terms also for the two crossed channels, with Mandelstam variables $u_E$ 
and $t_E$, which we denote by $C_2$ and $C_3$. Thus,
\begin{equation} \label{gamma4-comp-pole}
         \Gamma_{4V}^{[r,0]}=-\big(C_1^{[r]}\big)_{2V,2V}-
      \big(C_2^{[r]}\big)_{2V,2V}-\big(C_3^{[r]}\big)_{2V,2V}+V_{4V}^{[r,0]},
\end{equation}
and the derivation shows that the $V_{4V}^{[r,0]}$ defined by this
relation contains no more nonperturbative denominator structure 
in $s_E$, $u_E$, or $t_E$. The latter conclusion, strictly speaking, 
follows only for the adjoint color representation in each 
two-body channel, since in Fig.\ \dsdreigluonfig\  the leftmost 
gluon line projects $T'_{4V}$ onto this color subspace. It requires 
additional considerations, based on the Bethe-Salpeter normalization 
conditions, to check for possible zeroth-order, Mandelstam-variable poles 
in the other colored channels. Since these considerations logically belong 
to the discussion of the four-gluon amplitude, we defer them to the companion 
article \cite{4}. Here we anticipate the result: {\it In zeroth perturbative 
order}, there are no Mandelstam-variable poles in the other color sectors. 
Thus the "reduced" vertex function $V_{4V}^{[r,0]}$ has nonperturbative 
structure, rationally approximated, only in the variables 
$p_1^2,\ldots p_4^2$ of individual external legs. We emphasize that this in 
no way precludes the existence of glueball-type bound states in color-singlet 
channels. Such bound states are not elements of the generalized Feynman rule 
$\Gamma_{4V}^{[r,0]}$ at any level $r$, but arise through the standard 
mechanism of partial ( e.g., ladder ) resummation of quasi-perturbative 
corrections $(g/4\pi)^{2p}\Gamma_{4V}^{[r,p)}$ to all orders $p\ge 1$. 

The poles of (\ref{t-comp-pole}) cannot represent bound states, since in 
general their residues are not positive definite ( at $r=1$, for example, 
$u_{1,3}$ will be found to be positive ). Thus for the moment they would seem 
to be unphysical artefacts. But one immediately realizes that in fact they 
play a legitimate role by cancelling another unphysical phenomenon. Inspect 
the analytic structure of the one-gluon-reducible terms $A_i$ at level 
$[r,0]\,$ ( i.e., with $[r,0]$ diagram elements ). Again suppressing all 
unnecessary arguments, one has
\begin{equation}
       A_1^{[r,0]}\,=\,\Gamma_{3V}^{[r,0]}\ D^{[r,0]}(P)\ \Gamma_{3V}^{[r,0]}.
\end{equation}
By the construction prescriptions for the extended Feynman rules, the gluon 
propagator $D^{[r,0]}$ contains, besides its $r+1$ poles, a product of r 
numerator zeroes of the form
\begin{equation}
           \prod_{n=1}^r(P^2+u_{r,2n}\Lambda^2).
\end{equation}
On the other hand, both $\Gamma_{3V}^{[r,0]}$ vertices contain the same 
product in their denominators, so that there remain on the internal gluon 
line, in addition to the legitimate poles from $D$ describing propagation of 
the exchanged object, a number $r$ of extra poles at positions 
$P^2=-u_{r,2n}\Lambda^2$. Such extra poles are unacceptable physically; they 
would imply that the generalized Feynman rule for $D$ is incomplete. Now 
isolate the $n$-th unphysical-pole piece of $A_1$. From 
(\ref{gamma-3v-partial}) and (\ref{prop-pole-sc}), it must involve 
$B_nt(P)B_n$; by computing the residue, the piece is found to be
\begin{equation}
B_n^{[r]}\ \frac{(u_{r,2n+1})^{-1}t(P)}{P^2+u_{r,2n}\Lambda^2}\ B_n^{[r]},
\end{equation}
so that the sum of the unphysical pieces is the {\it negative} of (\ref{2v2v-comp-pole}):
\begin{equation}
\sum_{n=1}^r(\mbox{ $n$-th unphysical pole of $A_1^{[r,0]}$ })=\big(C_1^{[r]}\big)_{2V,2V}.
\end{equation}
Upon combining (\ref{t4}) and (\ref{gamma4-comp-pole}) into
\begin{equation} \label{t4-decomp}
T_{4V}^{[r,0]}={A'}_1^{[r,0]}+{A'}_2^{[r,0]}+{A'}_3^{[r,0]}+V_{4V}^{[r,0]},
\end{equation}
the unphysical artefacts then cancel exactly to leave the "softened" exchange 
graphs
\begin{equation} \label{a-primed}
{A'}_i^{[r,0]}=A_i^{[r,0]}-\big(C_i^{[r]}\big)_{2V,2V}\quad (i=1,2,3).
\end{equation}
The $\Gamma_{4V}^{[r,0]}$ poles inferred through the $2V$ and $3V$ equations therefore turn out to be "compensating poles", cancelling unphysical parts in the one-gluon-reducible terms of $T_{4V}$. It is clear that in the context of the extended iterative scheme it is the artefact-free $A_i'$, rather than the original $A_i$, that represent the {\it physical} one-gluon exchange mechanism, and the artefact-free $V_{4V}^{[r,0]}$, rather than $\Gamma_{4V}^{[r,0]}$, that constitutes the physical extended Feynman rule for four-gluon interaction. On the other hand, in the quantity (\ref{t4-prime}),
\begin{equation} \label{t4-prime-decomp}
{T'}_{4V}^{[r,0]}={A'}_2^{[r,0]}+{A'}_3^{[r,0]}+V_{4V}^{[r,0]}-\big(C_1^{[r]}\big)_{2V,2V},
\end{equation}
an uncompensated $C_1$ remains, enabling, as we have seen, the $(A)_3$ term of Fig.\ \dsdreigluonfig\  to produce the $B_n$ terms of (\ref{gamma-3v-partial}).

\subsection{Extended irreducibility. Rearranged vertex equation} 

In the general case where terms $(B)_3$, $(E)_3$, and (for $l\ge 2$ loops) $(D)_3$ contribute, one first generalizes (\ref{t-comp-pole}) to the conclusion that the $T$ matrices $T'_{G\bar G VV}$, $T'_{F\bar FVV}$, $T'_{5V}$ in general have poles (forming discrete approximations of cuts) at the same positions, $p_2^2=-u_{r,2n}\Lambda^2$, in their horizontal channels. One then invokes another general property of correlation functions \cite{8}: when a pole is present in several functions simultaneously, each residue factor is {\it uniquely} associated with its own subset of external lines, and independent of the remaining legs in the various functions. Thus at the pole $p_2^2=-u_{r,2n}\Lambda^2$, the residue factor for the rightmost two-gluon configuration in Fig.\ \dsdreigluonfig\  is the same $\Psi_n$ as in (\ref{t-comp-pole}) for all $T'$ amplitudes. Then (\ref{psi}) still follows, but (\ref{vert-pole-sc}) now has four different contributions on its l.h.s., corresponding to various dressing mechanisms for the gluon self-energy. Self-reproduction of the $B_n$ terms with $n\ge 1$ in (\ref{gamma-3v-partial}) is possible if $M_n$ and the various matrices taking the place of the $M_n^T$ in (\ref{vert-pole-sc}) are all the same mutiple of $t(p_2)$. Then the full two-gluon self-consistency conditions \cite{3} can be invoked, which generalize (\ref{prop-pole-sc}), and the $B_n$ term reproduces itself if (\ref{m-def}) is imposed.

As a by-product, one finds that the 1PI functions $\Gamma_{G\bar GVV}$, $\Gamma_{F\bar FVV}$ must contain terms of the form
\begin{align}
-\big(C_1^{[r]}\big)_{G\bar G,VV} & = -\sum_{n=1}^r{\tilde B}_n^{[r]}\,
\frac{(u_{r,2n+1})^{-1}t(P)}{P^2+u_{r,2n}\Lambda^2}\,B_n^{[r]}, 
                                                           \label{c1-ghost} \\
-\big(C_1^{[r]}\big)_{F\bar F,VV} & = -\sum_{n=1}^r{\bar B}_{F,n}^{[r]}\,
\frac{(u_{r,2n+1})^{-1}t(P)}{P^2+u_{r,2n}\Lambda^2}\,B_n^{[r]}, 
                                                               \label{c1-ferm}
\end{align}
in their two-body channels ($G+\bar G\leftrightarrow V+V$) and ($F+\bar F\leftrightarrow V+V$) respectively, where ${\tilde B}_n$, ${\bar B}_{F,n}$ are the amplitudes analogous to the $B_n$ of (\ref{gamma-3v-partial}) in the partial-fraction decompositions of the ghost vertex $\Gamma_{GV\bar G}^{[r,0]}(-q',k,q)$ and fermion vertices $\Gamma_{FV\bar F}^{[r,0]}(-p',k,p)$ with respect to their gluon-leg variable, $k^2$. The analogous but richer structure in $T'_{5V}$ will
be discussed in detail in \cite{4}.

The structure revealed by these residue-taking operations may look involved at first, but the final result is simple: the full 4-gluon, off-shell $T$ matrix (\ref{t4}), for example, has no unphysical artefacts at all. The artefacts arose because, in a nonperturbative context, the usual decomposition of $T$ by the criterion of ordinary one-particle (here, one-gluon) reducibility turns out to be an awkward one: both parts in such a division contain unphysical-pole terms that cancel in the sum. It is clearly more natural, and better suited to the physics of the problem, to perform the decomposition as in (\ref{t4-decomp}), where all parts are free of artefacts. To characterize such a decomposition more formally, we call the set of $r$ pole factors common to expressions (\ref{2v2v-comp-pole}) and (\ref{c1-ghost}/\ref{c1-ferm}) a gluonic shadow, described graphically by the double wiggly line of Fig.\ \softfig, and define as one-shadow-irreducible any amplitude built from $[r,0]$ extended Feynman rules that does not fall into two disconnected pieces upon cutting such a shadow line. The defining property of decomposition (\ref{t4-decomp}) then is that {\it all} its terms are one-shadow irreducible. In particular, $V_{4V}^{[r,0]}$ exhibits what one may call {\it extended irreducibility}, being irreducible both for gluon-propagator poles and for shadow poles, while the "softened" exchange diagrams ${A'}_i^{[r,0]}$, described graphically by using dotted diagram elements as in Fig.\ \softfig, are still reducible for the gluon-propagator-poles.

Use of (\ref{t4-prime-decomp}) and of the analogous decompositions
\begin{align}
{T'}_{G\bar GVV}^{[r,0]} & = \tilde A_2^{'[r,0]}+\tilde A_3^{'[r,0]}
+V_{G\bar GVV}^{[r,0]}-\big(C_1^{[r]}\big)_{G\bar G,VV}, \\
{T'}_{F\bar FVV}^{[r,0]} & = \bar A_{F,2}^{'[r,0]}+\bar A_{F,3}^{'[r,0]}
+V_{F\bar FVV}^{[r,0]}-\big(C_1^{[r]}\big)_{F\bar F,VV},
\end{align}
where ${\tilde A}'_i$ and ${\bar A}'_{F,i}$ have obvious meanings as softened exchange diagrams reducible for ghost or fermion propagator poles but with ghost-shadow (in non-Landau gauges only) or quark-shadow poles compensated, now leads to the rearranged vertex equation of Fig.\ \dsbnullgluonfig. With the pole terms of (\ref{gamma-3v-partial}) having been reproduced on the r.h.s.\ through condition (\ref{m-def}) on the {\it four}-gluon function, and with the second line of (\ref{gamma-3v-partial}) anticipated to be absent by restriction (\ref{gamma3-pc}) below, only an equation for the amplitude $B_0^{[r]}$ -- an object with three-gluon tensor structure but only two scalar variables -- remains. This rather strong reduction of the original equation, due to the "all-in-one-blow" self-reproduction of the $B_n$ parts with $n\ge1$ through the presence of the compensating poles, generally causes a loss of self-consistency conditions and therefore {\it underdetermination}, which tends to counteract the overdetermination coming from the lack of manifest symmetry. (At $r=1$, for example, the $x_{1,5}$ coefficient of (\ref{f0}) appears only in $B_1^{[1]}$ and not in $B_0^{[1]}$, and therefore gets no self-consistency condition of its own).

In Fig.\ \dsbnullgluonfig, only the terms that can contribute to the self-reproduction of the extended Feynman rule (i.e., produce terms of zeroth perturbative order) at one loop have been made explicit. Thus diagrams containing $V_{G\bar GVV}$ or $V_{F\bar FVV}$ no more appear: for {\it these} amplitudes, which 
are superficially convergent, it is now true that after extraction of the shadow-pole terms they consist only of superficially convergent loops. We see here
that the argument of \cite{3} concerning these higher amplitudes needs a subtle
qualification: that argument did not take into account the possibility of
certain {\it treelike} structures, of zeroth perturbative order, which nevertheless appear in the 1PI functions. The shadow-pole terms are precisely such
structures. Yet their presence does {\it not} imply a proliferation of Feynman
rules, since they consist entirely of building blocks determined already at the
level of the basic vertices. For the $V$ amplitudes, obtained by subtracting these, the argument of \cite{3} goes through : their insertion into terms $(B)_3$ and $(E)_3$ of Fig.\ \dsdreigluonfig\   produces integrals for which the number of $\frac{1}{\epsilon}(\Lambda_\epsilon^2/\nu_0^2)^{-\epsilon}$ factors lags behind the number of $g_0^2$ prefactors by at least one, and which therefore give only quasi-perturbative corrections of order $p\ge 1$. Note that this would not have been true for the corresponding $\Gamma$ amplitudes before extraction of the nonperturbative shadow pieces. Without this extraction one would have missed the dot modifications, and therefore left unphysical artefacts, in the triangle diagrams $(A)_{3'}$, $(E,E')_{3'}$, and $(F,F')_{3'}$ of Fig.\ \dsbnullgluonfig.

For the self-consistency problem, the rearrangement for shadow irreducibility brings both simplifications and complications. On the one hand we must now require that the $A'_i$ of (\ref{a-primed}), when used as diagram-building blocks, preserve perturbative power counting, as did the original $A_i$ by construction. It is easy to check from (\ref{2v2v-comp-pole}) that this allows no terms 
with net positive powers of $p_3^2$ or $p_1^2$ in the $B_n$, or equivalently, 
none of the terms in the second line of the p.f.\ decomposition 
(\ref{gamma-3v-partial}). In the numerator polynomial for the complete 
$\Gamma_{3V}^{[r,0]}$ rational approximant, we therefore have powers
\begin{equation} \label{gamma3-pc}
(p_1^2)^{m_1}(p_2^2)^{m_2}(p_3^2)^{m_3}\quad\mbox{restricted by }m_{1,2,3}
       \le r,
\end{equation}
a restriction significantly stronger than the $m_i+m_j\le 2r$ $(i\neq j)$ inferred previously \cite{3}, and which corresponds to restricting the extra polynomial $p_i^2$ dependence of the spectral functions in (\ref{spectral}) to at most a bilinear one. At $r=1$, in particular, this restriction forces the primed coefficients of (\ref{f0}) to zero:
\begin{equation}
x'_{1,2}=0\quad;\quad x'_{1,4}=0.
\end{equation}
Moreover it simplifies the writing of the softened one-gluon exchange mechanisms (\ref{a-primed}), since these can now be obtained simply by enumerating the physical gluon-propagator poles with their residues: in Landau gauge fixing,
\begin{eqnarray}
{A'}_1^{[r,0]}(k_1,\ldots k_4)=                          \nonumber \\
\sum_{m=0}^r \big[\Gamma_{3V}(k_1,k_2,P)\big]_{P^2=-\sigma_{r,2m+1}\Lambda^2}\frac{\rho_m t(P)}{P^2+\sigma_{r,2m+1}\Lambda^2}\big[\Gamma_{3V}(P,k_3,k_4)\big]_{P^2=-\sigma_{r,2m+1}\Lambda^2},
\end{eqnarray}
where
\begin{equation}
\rho_m=
\big[(P^2+\sigma_{r,2m+1}\Lambda^2)D_T(P)\big]_{P^2=-\sigma_{r,2m+1}\Lambda^2}
=\frac{\prod\limits_{n=1}^r(u_{r,2n}-\sigma_{r,2m+1})}{\prod\limits_{\atfrac{n=0}{(n\neq m)}}^r(\sigma_{r,2n+1}-\sigma_{r,2m+1})},
\end{equation}
and where $P=k_1+k_2=-(k_3+k_4)$. In non-Landau gauges, of course, the longitudinal-gluon propagator term of the original $A_1$ must be added unchanged.

On the other hand the rearrangement leads to the result (which is true generally but was not yet visible in the simpler case of the two-point equation at one loop treated in \cite{3}) that the perturbative limit as $\Lambda\to 0$ cannot, at low levels $r$, be maintained exactly in all amplitudes, but only asymptotically for increasing $r$. For $\Lambda\to 0$, the $V_{4V}^{[r,0]}$ of (\ref{gamma4-comp-pole}), as well as the $A_i^{[r,0]}$ of (\ref{a-primed}), go over into their zeroth-order perturbative counterparts, but the $C_i^{[r]}$ for low $r$ do not go to zero:
\begin{equation} \label{c1-pert}
\big[\big(C_1^{[r]}\big)_{2V,2V}\big]_{\Lambda=0}=\frac{1}{P^2}\Big\{\sum_{n=1}^rB_n^{[r]}(\Lambda=0)\frac{t(P)}{u_{r,2n+1}}B_n^{[r]}(\Lambda=0)\Big\}.
\end{equation}
Thus diagrams involving the one-gluon-exchange mechanisms $A_i'$, $\tilde A_i'$, $\bar A_{F,i}'$, such as the triangle diagrams of Fig.\ \dsbnullgluonfig, are expected not to exhibit a fully correct perturbative limit as $\Lambda\to 0$ at low levels $r$. The condition (or rather conditions, because there are several tensor structures involved) for the curly bracket in (\ref{c1-pert}) to vanish, of which we will present examples, can at best be fulfilled asymptotically for large $r$, where they spread their restrictive effect over an increasing number of nonperturbative vertex coefficients, and thus become progressively easier to maintain.

\subsection{3-gluon, $r=1$ self-consistency conditions} 

To extract the self-reproduction conditions of the extended Feynman rule $\Gamma_{3T}^{[1,0]}$ one evaluates, with eqs.\ (\ref{pi}/\ref{pi-eps}) in mind, the divergent parts of the terms on the r.h.s.\ of Fig.\ \dsbnullgluonfig\  in dimensional regularization and Landau gauge fixing, with $[1,0]$ diagram elements throughout. The restriction to divergent parts makes these calculations 
somewhat analogous to the computation of one-loop renormalization constants in perturbation theory (and much more feasible than full evaluations of $[1,1)$ radiative corrections, which already at $r=1$ are very lengthy). For the input $V_{4T}^{[1,0]}$ to diagram $(C)_{3'}$, we anticipate formulas from appendix B 
of 
the companion article \cite{4}: this approximant represents a theoretically motivated restriction to a subset of fifteen of the many possible color and Lorentz tensor structures of a four-gluon amplitude, with invariant functions characterized by a set $\zeta=\{\zeta_1,\ldots\zeta_{17}\}$ of seventeen dimensionless, real numerator coefficients. The terms proportional to the number $N_F$ of quark flavors, arising from fermion-loop diagrams $(F)_{3'}$ and $(F')_{3'}$, are valid for {\it massless quarks} ($\hat m_F=0$ in the notation of the appendix of \cite{3}), where all fermionic mass scales, too, are simply multiples of $\Lambda$. For brevity, we abstain from listing contributions of the various diagrams separately \cite{9} and present only the combined results. Eqs.\ (\ref{num-sc}) for the coefficients $x_{1,i}$ (now written $x_i$ for brevity) of the $F_0^{[1,0]}$ invariant functions (\ref{f0}) are:
\begin{multline}
\hspace{1.5cm}\frac{1}{\beta_0}\Big[ 
-\frac{9}{4}x_1+\frac{15}{16}x_3+\frac{1}{u_3}\Big(\frac{1}{4}x_1x_2-9x_1x_4+x_3x_4\Big) \\
+\frac{2}{3}N_F\Big(z_3+\frac{1}{u_3}x_4z_3-\frac{1}{w_3}z_1z_4\Big)\Big] =x_1\hspace{1.5cm} \label{sce-3gl-1}
\end{multline}
\begin{multline}
\hspace{1.5cm}\frac{1}{\beta_0}\Big[ 
\frac{3}{2}x_3^2+\frac{1}{u_3}\Big(\frac{1}{2}x_2x_4-2x_4^2-\frac{15}{2}x_1x_5+\frac{5}{4}x_3x_5\Big) \\
+\frac{2}{3}N_F\Big(z_3^2+\frac{1}{u_3}x_5z_3-\frac{1}{w_3}z_4^2\Big)\Big] =x_2\hspace{1.5cm} \label{sce-3gl-2}
\end{multline}
\begin{multline}
\hspace{1.5cm}\frac{1}{\beta_0}\Big[ 
\frac{3}{2}x_3+\frac{1}{u_3}\Big(-\frac{37}{4}x_1x_4+\frac{3}{2}x_3x_4\Big)-Z_1(\zeta) \\
+\frac{2}{3}N_F\Big(z_3+\frac{1}{u_3}x_4z_3-\frac{1}{w_3}z_1z_4\Big)\Big] =x_3\hspace{1.5cm} \label{sce-3gl-3}
\end{multline}
\begin{multline}
\hspace{1.5cm}\frac{1}{\beta_0}\Big[ 
-\frac{9}{4}x_1+\frac{15}{16}x_3+\frac{1}{u_3}\Big(-\frac{31}{4}x_1x_2-\frac{5}{4}x_1x_4+\frac{5}{4}x_2x_3\Big)-Z_1(\zeta) \\
+\frac{2}{3}N_F\Big(z_3+\frac{1}{u_3}x_2z_3-\frac{1}{w_3}z_1z_4\Big)\Big] =x_1\hspace{1.5cm} \label{sce-3gl-4}
\end{multline}
\begin{multline}
\hspace{1.5cm}\frac{1}{\beta_0}\Big[ 
\frac{3}{2}x_3^2+\frac{1}{u_3}\Big(-\frac{1}{4}x_2x_4-\frac{5}{4}x_4^2-\frac{15}{2}x_1x_5+\frac{5}{4}x_3x_5\Big)-Z_2(\zeta) \\
+\frac{2}{3}N_F\Big(z_{3}^2+\frac{1}{u_3}x_5z_{3}-\frac{1}{w_3}z_{4}^2\Big)\Big] =x_4\hspace{1.5cm} \label{sce-3gl-5}
\end{multline}
(The fermionic vertex-coefficients $z_i$ are defined by eq.\ (\ref{ferm-coef}) below.)

On the other hand, the terms of the invariant function $F_1$ of (\ref{f1}) turn out to be fed only by themselves, and by terms in $V_{4V}$ which our above-mentioned, restricted form of that amplitude omits in the first place. Again, since we view the basic vertices as an interrelated whole, it did not seem consistent to us to keep just one source of such terms. Thus
\begin{equation}
x_6=0\quad;\quad x_7=0 
\end{equation}
is a consistent and {\it self-consistent} choice in our framework.

A noteworthy feature is that the coupling to the 4-gluon amplitude enters 
only into the three equations\ (\ref{sce-3gl-3}-\ref{sce-3gl-5}), and only 
through two linear combinations of its seventeen coefficients $\zeta$,
\begin{align}
Z_1(\zeta) & = \frac{15}{32}\,(\,3\zeta_{1}-\zeta_{7}\,),      \label{z1} \\
Z_2(\zeta) & = \frac{15}{32}\,(\,3\zeta_{2}+3\zeta_{3}
              -\zeta_{8}-\zeta_{9}\,)\,.                       \label{z2}
\end{align}
The observation that the rather large (and, as it will turn out, strongly overdetermined) self-consistency problem of the four-point vertex parameters couples to the 2-point and 3-point problem only through this narrow "bottleneck" will be important as it will suggest ways of breaking down the rather voluminous total self-consistency problem into more manageable pieces.

As already noted, there is no equation with $x_5$ on its r.h.s. But $B_0^{[1]}$ still has three-gluon tensor structure, and $x_1$, by (\ref{gl-3-vert}), appears twice in conjunction with two different tensor structures, so there are {\it two} equations, (\ref{sce-3gl-1}) and (\ref{sce-3gl-4}), for $x_1$. The relation obtained by subtracting these,
\begin{equation}
\frac{1}{u_3}\Big(8x_1x_2-\frac{31}{4}x_1x_4+x_3x_4-\frac{5}{4}x_2x_3\Big)+Z_1(\zeta)-\frac{2}{3}N_F\Big(\frac{x_2-x_4}{u_3}\Big)z_3=0,
\end{equation}
represents the imposition, in zeroth perturbative order, of Bose symmetry on a DS equation that is not manifestly Bose symmetric. It appears to be fortuitous that the "loss" of one equation, incurred in the reduction of the self-consistency problem to the partial amplitude $B_0^{[1]}$, is just compensated by one Bose-symmetry restriction; we are not aware of a deeper reason for this phenomenon.

Finally we note the result for the coefficient of the perturbative-remainder divergence, the $\Xi_N^{(1)}$ of eq.\ (\ref{functional-iter}):
\begin{align}
\Big(\Xi_{3T}^{(1)}\Big)^{\rho\kappa\sigma} = 
\quad  & \delta^{\kappa\sigma}(p_2-p_3)^\rho \Big[-\frac{17}{4}+\frac{1}{u_3}\Big(-x_1^2-8x_1x_3+\frac{5}{4}x_3^2\Big)+\frac{2}{3}N_F\Big(1+\frac{1}{u_3}x_3z_3-\frac{1}{w_3}z_1^2\Big)\Big] \nonumber \\ 
 + \, & \delta^{\sigma\rho}(p_3-p_1)^\kappa \Big[-\frac{17}{4}+\frac{1}{u_3}\Big(-\frac{37}{4}x_1^2+\frac{3}{2}x_1x_3\Big)+\frac{2}{3}N_F\Big(1+\frac{1}{u_3}x_1z_3-\frac{1}{w_3}z_1^2\Big)\Big] \nonumber \\ 
 + \, & \delta^{\rho\kappa}(p_1-p_2)^\sigma \Big[-\frac{17}{4}+\frac{1}{u_3}\Big(-\frac{37}{4}x_1^2+\frac{3}{2}x_1x_3\Big)+\frac{2}{3}N_F\Big(1+\frac{1}{u_3}x_1z_3-\frac{1}{w_3}z_1^2\Big)\Big] \label{3g-pert-defect}
\end{align}
The purely perturbative result would be of the form (\ref{pert-sc}) with the Landau-gauge value
\begin{equation}
z_{3V}^{(1)}(\xi=0)=-\frac{17}{4}+\frac{2}{3}N_F.
\end{equation}
The existence, and lack of complete Bose symmetry, of deviations from the perfect perturbative limit do not come as a surprise in view of what we noted before in connection with (\ref{c1-pert}). Eliminating them would again produce overdetermination of the $x$ coefficients and generally is not feasible exactly for low $r$ but only asymptotically for large $r$.

\section{Equations for the ghost vertices}
\setcounter{equation}{0}

For the ghost-gluon-antighost vertex, $\Gamma_{GV\bar G}$, and its generalized Feynman rule, it is again self-consistent (though not the most general solution) to assume a pure $f_{abc}$ color structure, the Lorentz structure then being given by 
\begin{equation} \label{gamma_ghost}
\big[\Gamma_{GV\bar G}^{[r,0]}(p,k,-p')\big]_{abc}^\mu
=if_{abc}\big[p^\mu\tilde F_0^{[r,0]}(p,k,-p')+k^\mu\tilde F_1^{[r,0]}(p,k,-p')\big]
\end{equation}
The dimensionless invariant function $\tilde F_i$, with perturbative limits $\tilde F_i^{(0)pert}=\delta_{i0}$, depend on the invariants $p^2$, $k^2$, ${p'}^2$. Fig.\ \dsdreighostfig\  shows the diagrammatic form of the DS equation for $\Gamma_{GV\bar G}$ in its ghost channel, i.e., with the "$G$" leg as the unsymmetrically distinguished leftmost leg. It again features a four-point amplitude $T_{GVV\bar G}'$ which is 1PI in only the "horizontal" channel. Residue-taking both in this equation and in the corresponding equation in the antighost channel again reveals the presence of compensating poles in $T'$, and taking these into account one again obtains a rearranged form of the equation as in Fig.\ \dsdreighostfig(b).

When staying strictly in Landau gauge, as we do in this paper, it is actually unnecessary, as far as the self-consistency of the generalized Feynman rule is concerned, to evaluate the terms on the r.h.s.\ of Fig.\ \dsdreighostfig(b) in detail: brief inspection shows that the latter, at $\xi=0$, do not sustain nonperturbative $\Lambda$ terms. Consider e.g., term $(B)_G$ of Fig.\ \dsdreighostfig(b) with the momentum assignments shown. Its upper $\Gamma_{GV\bar G}$ vertex has, by (\ref{gamma_ghost}), Lorentz structure
\begin{equation}
(p'-q_2)^\lambda\tilde F_0+q_2^\lambda\tilde F_1.
\end{equation}
The internal gluon line carrying momentum $q_2$ has, in Landau gauge, a transverse projector $t^{\kappa\lambda}(q_2)$ so that only the ${p'}^\lambda\tilde F_0$ portion survives. But $p'$ is an external momentum not running in the loop, so the integrand loses one power of the loop momentum as compared to standard power counting. Since the loop had only logarithmic divergence to begin with, it is now actually convergent. Analogous arguments apply to the term $(C)_G$ of Fig.\ \dsdreighostfig(b). The only possible nonperturbative modifications are therefore those from term $(A)_G$, if any. These must have at least one denominator factor $\big(p^2+\tilde u_{1,2}'\Lambda^2\big)$ in the momentum variable of the leftmost (ghost) leg. But the same argument applies to the two alternative forms of the ghost-vertex equation not displayed in Fig.\ \dsdreighostfig, with the "antighost" (momentum $-p'$) and gluon (momentum $k$) lines, respectively, as leftmost legs. The only nonperturbative terms in the $\tilde F$ invariant functions of equation (\ref{gamma_ghost}), that are candidates for self-consistency, are therefore those proportional to
\[
\frac{\Lambda^6}{\big(p^2+\tilde u_{1,2}'\Lambda^2\big)
                 \big(k^2+u_{1,2}'\Lambda^2\big)
                 \big({p'}^2+\tilde u_{1,2}'\Lambda^2\big)}.
\] 
These terms, however, make the loop appearing in term $(A)_G$ of Fig.\ \dsdreighostfig(b) convergent. Since the self-reproduction mechanism of eqs.\ (\ref{pi}/\ref{pi-eps}) is dependent upon the $\frac{1}{\epsilon}$ divergence factor, the amplitude cannot develop any zeroth-order nonperturbative terms in Landau gauge:
\begin{equation} \label{gamma_ghost_pert}
\big[\Gamma_{GV\bar G}^{[r,0]}(p,k,-p')\big]_{abc}^\mu= if_{abc}p^\mu
= \big(\Gamma_{GV\bar G}^{(0)pert}\big)_{abc}^\mu
\qquad (\xi=0,\mbox{ all }r). 
\end{equation}
The effect is, of course, basically familiar from perturbation theory: there, the one-loop divergence of the renormalization constant $\tilde Z_1$ vanishes at $\xi=0$. The preceding discussion merely serves as a reminder that in the present context such special divergence reductions also have qualitative dynamical consequences as they suppresss the divergence-related self-consistency mechanism. For an amplitude with unphysical degrees of freedom such as $\Gamma_{GV\bar G}$, it is of course legitimate to depend on the gauge fixing in this way.

The ghost-self-energy equation, due to its general divergence reduction as discussed in appendix A.2 of \cite{3}, also has effectively a logarithmically divergent integral. When evaluated with the purely perturbative vertex (\ref{gamma_ghost_pert}), that integral can produce no more than the {\it perturbative} divergence, so that again no nonperturbative terms are formed:
\begin{equation}
\tilde D^{[r,0]}(p^2)=\tilde D^{(0)pert}(p^2)
=\frac{1}{p^2} \qquad (\xi=0,\mbox{ all }r).
\end{equation}
We see that the assumption of refs.\ \cite{5} that ghost vertices remain perturbative is justified only in Landau gauge.

\section{Massless-fermion vertices}
\setcounter{equation}{0}

In the absence of Lagrangian mass terms for quarks, nonperturbative mass 
scales in the two basic fermion vertices $\Gamma_{F\bar F}$ and 
$\Gamma_{FV\bar F}$ can only be multiples of the $\Lambda$ scale. This case 
is technically far simpler than the rather complicated situation encountered 
in the presence of additional RG-invariant scales from "current" quark 
masses, and is the only one we consider in this paper.

The DS equation for the inverse quark propagator is given diagrammatically 
in Fig.\ \dszweifermfig. The corresponding extended Feynman rule at level 
$r=1$, as discussed in the appendix of \cite{3}, is 
\begin{equation}
       -\Gamma_{F\bar F}^{[1,0]}(p)=\slash p+w_1\Lambda+\frac{w_3\Lambda^2}
              {\slash p+w_2\Lambda}. \label{ferm-prop}
\end{equation}
Its nonperturbative content is characterized by the three dimensionless, real 
parameters $w_1$, $w_2$, $w_3$. Extraction of the self-consistency conditions 
for these involves calculating the divergent parts of the loop of Fig.\ 
\dszweifermfig, evaluated with $[1,0]$ input elements, and proceeds largely 
as in the gluon-propagator case considered in \cite{3}. The resulting 
equations
\begin{align}
         w_2 & = w_2'\quad, \label{ferm_pole_parameter} \\
      \frac{1}{\beta_0}\Big[4w_1-4z_1\Big] & =w_1\quad, \label{sce-2fe-1}  \\
      \frac{1}{\beta_0}\Big[4w_1z_1-4z_2\Big] & =w_3 \quad, \label{sce-2fe-2} 
\end{align} 
make reference to the parameters $z$ of the quark-gluon three-point vertex, 
$\Gamma_{FV\bar F}$. Its transverse-gluon projection at level $r=1$ reads,
\begin{multline}
      \big[\Gamma_{FT\bar F}^{[1,0]}(-p',k,p)\big]^\nu
                       = \,t^{\nu\mu}(k)\bigg\{\gamma^\mu 
                +z_1\Big(\frac{\Lambda}{\slash p'+w_2'\Lambda}\gamma^\mu
                +\gamma^\mu\frac{\Lambda}{\slash p+w_2'\Lambda}\Big)      \\
                +z_2\frac{\Lambda}{\slash p'+w_2'\Lambda}\gamma^\mu
                        \frac{\Lambda}{\slash p+w_2'\Lambda} 
                +\frac{\Lambda^2}{k^2+\bar u_2'\Lambda^2} \Big[z_3\gamma^\mu
                +z_4\Big(\frac{\Lambda}{\slash p'+w_2'\Lambda}\gamma^\mu
                +\gamma^\mu\frac{\Lambda}{\slash p+w_2'\Lambda}\Big)      \\
                +z_5\frac{\Lambda}{\slash p'+w_2'\Lambda}\gamma^\mu
            \frac{\Lambda}{\slash p+w_2'\Lambda} \Big]\bigg\}.
\end{multline}
Here the notation of ref.\ \cite{3} for the dimensionless coefficients has 
been changed and simplified somewhat, the relation \cite{3} $\to$ this paper 
being given by 
\begin{equation}
        z_{0,1}^{[1]}\to z_1,\quad z_{0,4}^{[1]}\to z_2,\quad 
        z_{1,0}^{[1]}\to z_3,\quad z_{1,1}^{[1]}\to z_4,\quad 
        z_{1,4}^{[1]}\to z_5. \label{ferm-coef}
\end{equation}
We have from the outset omitted all terms that would lead to conflict with 
perturbative divergence degrees. For $\Gamma_{FV\bar F}$ there are two DS 
equations, one in the "fermionic" and one in the "gluonic" channel, which 
are equivalent for the exact vertex but in general will give rise to 
different approximations. Rearranged for one-quark-shadow irreducibility in 
the now familiar way, these are depicted in Figs.\ \dsdreifermfig(a) and 
\dsdreifermfig(b), respectively. While each of the two forms, due to the 
compensating-poles mechanism, suffers from "loss of equations" in the sense 
discussed in sect.\ 2.3, it is interesting that {\it the two forms taken 
together} produce just the required number of self-consistency conditions:
\begin{align}
\frac{1}{\beta_0}\Big[
\frac{9}{4}z_1-\frac{9}{4}\frac{1}{w_3}z_1z_2
-\frac{1}{u_3}\Big(\frac{15}{2}x_1-\frac{5}{4}x_3\Big)z_4
+\frac{2}{3}N_F\frac{1}{u_3}z_3z_4\Big]  & = z_1 \label{sce-3fe-1} \\
\frac{1}{\beta_0}\Big[
\frac{9}{4}z_1^2-\frac{9}{4}\frac{1}{w_3}z_2^2
-\frac{1}{u_3}\Big(\frac{15}{2}x_1-\frac{5}{4}x_3\Big)z_5
+\frac{2}{3}N_F\frac{1}{u_3}z_3z_5\Big]  & = z_2 \label{sce-3fe-2} \\
\frac{1}{\beta_0}\Big[
\frac{9}{4}z_1-\frac{9}{4}\frac{1}{u_3}x_1z_4 \Big]& = z_1 
                                                 \label{sce-3fe-3} \\ 
\frac{1}{\beta_0}\Big[
\frac{9}{4}x_3-\frac{9}{4}\frac{1}{u_3}x_4z_3 \Big]& = z_3 
                                                  \label{sce-3fe-4} \\ 
\frac{1}{\beta_0}\Big[
\frac{9}{4}x_3z_1-\frac{9}{4}\frac{1}{u_3}x_4z_4 \Big]& = z_4 
                                                    \label{sce-3fe-5}
\end{align}
In addition, the quasi-perturbative remainders contain divergences given by
\begin{align}
\Big(\Xi_{FT\bar F}^{(1)}\Big)^\mu=\gamma^\mu &
\Big[\frac{9}{4}-\frac{9}{4}\frac{1}{w_3}z_1^2
-\frac{1}{u_3}\Big(\frac{15}{2}x_1-\frac{5}{4}x_3\Big)z_3
+\frac{2}{3}N_F\frac{1}{u_3}z_3^2\Big], \label{3f-pert-defect-g} \\
\Big(\Xi_{FT\bar F}^{(1)}\Big)^\mu=\gamma^\mu &
\Big[\frac{9}{4}-\frac{9}{4}\frac{1}{u_3}x_1z_3\Big], 
                                          \label{3f-pert-defect-f}
\end{align}
for Figs. \dsdreifermfig(a) and \dsdreifermfig(b) respectively, which differ 
from the perturbative quantity
\begin{equation}
z_{FV\bar F}^{(1)}(\xi=0)
=\frac{9}{4}
\end{equation}
by defect terms involving the vertex constants $x$ and $z$, which again 
cannot be forced to zero at $r=1$ but only asymptotically at large $r$.

Note that there is no equation determining the vertex-pole position $w_2'$, 
which by (\ref{ferm_pole_parameter}) is also the propagator-zero position.

\section{Solution for 2- and 3-point coefficients}
\subsection{Analysis of equations}
\setcounter{equation}{0}

The system of self-consistency conditions for the $r=1$ nonperturbative 
coefficients, as established up to now, consists of eqs. (4.8/4.30/4.31)
of ref. \cite{3} for the gluonic self-energy parameters $u_1, u_2, u_3$,
which now assume\footnote{Eq. (\ref{u1cond}) corrects for a misprint in 
eq. (4.31) of the first of refs. \cite{3}, where a tadpole contribution 
$-9N_Cu_{1,1}/4$ appears with the wrong sign} the simpler forms
\begin{equation}
      u_2' \,=\, \bar{u}_2' \,=\, u_2\; ,               \label{u2cond}    
\end{equation}
\begin{equation}
      \frac{1}{\beta_0} \left[ \frac{9}{4}u_1-\frac{33}{2}x_1
           +\frac{5}{4}x_3-2N_F\bigg( w_3+(w_1+w_2)z_1+z_2-\frac{1}{3}
               z_3 \bigg) \right] \,=\, u_1\; ,         \label{u1cond}
\end{equation}
\begin{equation}
      \frac{1}{\beta_0} \left[ \frac{5}{2}u_2 \Big( 3x_1-\frac{1}{2}x_3
          \Big) +9u_1x_3-9x_4-2N_F\bigg( z_3\Big( w_3+\frac{1}{2}u_2 \Big)
            +(w_2-w_1)z_4+z_5 \bigg) \right] \,=\, u_3\; , \label{u3cond}
\end{equation}
plus eqs.\ (\ref{sce-3gl-1}-\ref{sce-3gl-5}) above for the 
3-gluon-coefficients $x_1...x_5$, plus eqs. 
(\ref{ferm_pole_parameter}-\ref{sce-2fe-2}) and 
(\ref{sce-3fe-1}-\ref{sce-3fe-5}) above for the self-energy 
coefficients $w_1...w_3$ and vertex coefficients $z_1...z_5$ of massless
fermions. Its peculiar properties, in particular with respect to under-
and overdetermination tendencies, can be summarized as follows.

{\bf (a) }
One observes that the system as a whole exhibits a {\it scaling property:} 
any one of the coefficients that is presumed to be nonzero may be divided 
out of these equations, while replacing the others by their {\it ratios} to 
this one or to a uniquely fixed power of it, and rescaling $\Lambda$ 
accordingly. This property, which in a nonlinear system is nontrivial, is a 
natural consequence of the scheme-blindness of the basic self-consistency 
mechanism: a rescaling of $\Lambda$, which corresponds to a change of scheme, 
will not change the form of the zeroth-order conditions. We will choose, for 
definiteness, a rescaling by the 3-gluon coefficient $x_1$ of (\ref{f0}):
\begin{equation}
\tilde\Lambda^2 = x_{1}\Lambda^2,                  \label{tilde-lambda}
\end{equation}
\begin{align}
\tilde u_{1} &= \frac{u_{1}}{x_{1}}, &
\tilde u_{2} &= \frac{u_{2}}{x_{1}}, &
\tilde u_{3} &= \frac{u_{3}}{x_{1}^2}, && && \nonumber \\
\tilde x_{1} &= 1, &
\tilde x_{2} &= \frac{x_{2}}{x_{1}^2}, &
\tilde x_{3} &= \frac{x_{3}}{x_{1}}, &
\tilde x_{4} &= \frac{x_{4}}{x_{1}^2}, &
\tilde x_{5} &= \frac{x_{5}}{x_{1}^3}, \nonumber \\
\tilde Z_1 &= \frac{Z_1}{x_{1}}, & 
\tilde Z_2 &= \frac{Z_2}{x_{1}^2}, && && && \nonumber\\
\tilde w_1 &= \frac{w_1}{\sqrt{x_1}}, &
\tilde w_2 &= \frac{w_2}{\sqrt{x_1}}, &
\tilde w_3 &= \frac{w_3}{x_1}, && && \nonumber \\
\tilde z_{1} &= \frac{z_{1}}{\sqrt{x_{1}}}, &
\tilde z_{2} &= \frac{z_{2}}{x_{1}}, &
\tilde z_{3} &= \frac{z_{3}}{x_{1}}, &
\tilde z_{4} &= \frac{z_{4}}{\sqrt{x_{1}}^3}, &
\tilde z_{5} &= \frac{z_{5}}{x_{1}^2}.             \label{tilde-coefs}
\end{align}
The reason for this choice is that by putting $x_{1}=0$ one would end up 
with only the trivial solution (all nonperturbative coefficients vanishing), 
so one is not losing interesting solutions by assuming $x_1\neq 0$.
( For the scaling of the fermionic parameters, we are for the moment 
assuming $x_{1}$ to be positive ).
In effect, the rescaling {\it reduces the number of unknowns by one} while 
introducing the modified $\tilde\Lambda$ of (\ref{tilde-lambda}), scaled by 
an unknown factor. 

The rescaled system of fourteen conditions, with fifteen
unknowns and $\tilde{Z}_1$, $\tilde{Z}_2$ as external parameters, has 
solutions coming in pairs: one checks that if
\begin{align}                                       \label{oricoef}
    \Big\{ \tilde{u}_1,\tilde{u}_2,\tilde{u}_3, \;\tilde{w}_1,\tilde{w}_2,
           \tilde{w}_3, \;\tilde{x}_2...\tilde{x}_5, \;\tilde{z}_1,
           \tilde{z}_2,\tilde{z}_3,\tilde{z}_4,\tilde{z}_5 \Big\}
\end{align}
is a solution, then for the same $\tilde{Z}_1$ and $\tilde{Z}_2$ the set
\begin{align}                                       \label{mirrorcoef}     
    \Big\{ \tilde{u}_1,\tilde{u}_2,\tilde{u}_3, \;-\tilde{w}_1,-\tilde{w}_2,
           \tilde{w}_3, \;\tilde{x}_2...\tilde{x}_5, \;-\tilde{z}_1,
           \tilde{z}_2,\tilde{z}_3,-\tilde{z}_4,\tilde{z}_5 \Big\}
\end{align} 
is also a solution, which we shall refer to as a "mirror" solution. Note
that this discrete ambiguity affects only fermionic parameters.

{\bf (b) }
We have not obtained equations fixing the vertex-denominator parameters 
($u_{2}$ and $w_{2}$ in the present case). Neither analysis of the 4-gluon 
vertex \cite{4} nor, as preliminary studies indicate, use of 
"resummed" DS equations will change this situation. We did obtain 
conditions like the $w_{2}'=w_{2}$ of eq.\ (\ref{ferm_pole_parameter}), 
and the corresponding $u_2'=\bar u_2'=u_2$ of eq.\ (\ref{u2cond}), 
which ensure one common pole position in all basic vertices 
for a given type of external leg (and also the presence of 
propagator zeroes at the positions of vertex poles), but $u_2$ and $w_2$, 
in the end, have no determining equations of their own. This leads to 
the unexpected conclusion that the divergent parts of the momentum-space
DS equations as used up to now do not yet determine
the nonperturbative $\Lambda$ dependence completely. The reason 
is that these equations, in a sense, do not provide enough divergence. 
Indeed, the quadratically divergent gluon self-energy is the only vertex 
having $u_2$ and $w_2$ appear at least on the {\em right-hand} sides of its 
self-consistency conditions for $u_1$ and $u_3$, but at least two more 
equations with the same or higher degree of divergence would be needed to
``lift'' the two parameters from the denominators of loop integrands
into numerator expressions that provide self-consistency conditions -- 
a feat that only divergent integrations can perform.

Additional conditions for fixing $u_2$ and $w_2$ therefore should 
have general compatibility with the momentum-space DS equations and 
provide sufficient divergence. The only natural candidates here are those
requiring the vanishing of the "equation-of-motion condensates", i.e. of 
vacuum expectations of the simplest (dimension four) local composite 
operators proportional to the left-hand sides of the field equations. We use 
the condensate conditions for the ghost and fermion fields in the form
\begin{align}
\left(g_0\nu_0^\epsilon\right)^2
\left<0\right|\bar c_a(x)\left\{\left[\delta_{ab}\Box
+g_0\nu_0^\epsilon f_{abc}A_c^\mu(x)\partial^\mu
\right]c_b(x)\right\}\left|0\right> &=0 \\
\left(g_0\nu_0^\epsilon\right)^2
\left<0\right|\bar \psi(x)\left\{\left[i\slash\partial+g_0\nu_0^\epsilon
\slash A(x)
\right]\psi(x)\right\}\left|0\right> &=0
\end{align}
In momentum space these are, of course, nothing but the ghost and quark 
propagator equations integrated over momentum space, or equivalently,
taken at zero separation in coordinate space. They are therefore
obviously compatible with, and natural completions of, the unintegrated
( momentum space ) or nonzero-separation ( coordinate space ) DS equations
we have exploited up to now. In standard integral-equation theory
with convergent integrals and well-behaved functions, they would not
represent independent statements, but in a theory with divergent loop
integrals and therefore in need of renormalization, 
they do carry new information: since they involve operators 
of higher compositeness, they possess new divergences leading to new 
{\it zeroth-order} conditions on the nonperturbative coefficients.
At the $r=1$ level these conditions read,
\begin{equation}
\frac{3}{\beta_0^2}\left[-u_1^2+u_3\right]\Lambda^4 =0     \label{cond-ghost}
\end{equation}
\begin{multline}
\bigg\{\frac{3}{\beta_0}\Big[
w_1^4-4w_1^2w_3-2w_1w_2w_3-w_2^2w_3+w_3^2
\Big]\\
\shoveleft{ +\frac{2}{\beta_0^2}\Big[
u_3-u_1^2-3u_1w_3-6w_1^4+12w_1^2w_3+6w_1w_2w_3 } \\
\shoveright{ 
+3z_1(u_1w_1-u_1w_2+4w_1^3+2w_1^2w_2+2w_1w_2^2-6w_1w_3-2w_2w_3) 
\hspace{1.5cm} } \\
\shoveright{
+3z_2(-u_1-2w_1^2-2w_1w_2-2w_2^2+2w_3)
+z_3 (u_1+u_2+3w_3)
+3z_4(w_2-w_1)+3z_5
\Big]\bigg\} \  \Lambda^4 =0  }                           \label{cond-quark}
\end{multline}

Condition (\ref{cond-ghost}), from the ghost equation of motion, notably 
provides a restriction on {\it gluonic} parameters ( it is, incidentally, 
equivalent to requiring the vanishing of the zeroth-order, dimension-two 
gluon condensate $\langle A^{\mu} A^{\mu} \rangle$, which at $r=1$
turns out to be proportional to $-u_1^{\,2}+u_3$ ). Since in Landau
gauge at one loop it is the only condition from the ghost sector, is
independent of the presence of fermions, and of remarkable simplicity,
we give it priority in complementing eqs.(\ref{u1cond}/\ref{u3cond}). 
Condition (\ref{cond-quark}) brings in, in addition, the fermionic 
parameters, and is suitable for complementing eqs.
(\ref{sce-2fe-1}/\ref{sce-2fe-2}). The order-$[1,0]$ 
equation-of-motion condensate for the gluon field,
\begin{multline}
          \left(g_0\nu_0^\epsilon\right)^2
               \big<0\big|\,\big[\frac{1}{2}\big(\partial^\mu 
                A^\nu_a-\partial^\nu A^\mu_a\big)^2+\frac{3}{2}
          \left(g_0\nu_0^\epsilon\right)f_{abc}\big(\partial^\mu 
                A^\nu_a-\partial^\nu A^\mu_a\big)A_b^\mu A_c^\nu \\
         +\left(g_0\nu_0^\epsilon\right)^2 f_{abe}f_{cde}A_a^\mu 
                  A_b^\nu A_c^\mu A_d^\nu
         +\left(g_0\nu_0^\epsilon\right) f_{abc}
                 \big(\partial^\mu\bar c_a\big) A_c^\mu c_b
         +\left(g_0\nu_0^\epsilon\right)N_F\bar 
                               \psi\slash A\psi\big]\,\big|0\big> =0,
\end{multline}
is the most complicated, and its zeroth-order form at $r=1$,
\begin{multline}
         \bigg\{\frac{24}{\beta_0}\Big[u_1^2-u_3\Big] \\
         \shoveleft{+\frac{1}{\beta_0^2}\Big[
          -390 u_1^2+228u_3-324x_4+594u_1x_1+279u_1x_3+270u_2x_1-45u_2x_3}
                \Big] \\
        \shoveleft{+\frac{N_F}{\beta_0^2}\Big[
          8u_1^2-8u_3+24u_1w_3+48w_1^2-96w_1^2w_3-48w_1w_2w_3 } \\
          \shoveright{+24z_1(-u_1w_1+u_1w_2-4w_1^3-2w_1^2w_2-
                   2w_1w_2^2+6w_1w_3+2w_2w_3)  \hspace{1.5cm} } \\
        \shoveright{+24z_2(u_1+2w_1^2+2w_1w_2+2w_2^2-2w_3)
                    -8z_3(u_1+u_2+3w_3)+24z_4(w_1-w_2)-24z_5\Big] 
                                   \hspace{1.5cm} } \\
        +\ \frac{1}{\beta_0^3}\ \Big[\quad \mbox{3-loop-terms}\quad \Big]
                              \bigg\}\Lambda^4=0,      \label{cond-gluon}
\end{multline}
involves the largest number of vertex parameters simultaneously. In the
present context it cannot be applied in its exact form; the 3-loop-terms
must be omitted for formal consistency with the omission of 2-loop-terms in
the $l=1$ gluon-self-energy calculation. For these reasons, condition
(\ref{cond-gluon}) may be expected to be the most difficult to fulfill
on our level of approximation, and is not one of our primary choices
for completing the self-consistency system. We will check in the end to
what extent it can be accommodated.

{\bf (c) }  
Due to the large dimensions and considerable overdetermination 
of the 4-gluon self-consistency problem to be discussed in \cite{4}, 
the total coupled problem cannot, in our experience, be attacked directly 
with currently existent mathematical software tools. However,
we have already emphasized the ( also unexpected ) result that the
4-gluon-vertex problem couples to the fewer-point amplitudes only through 
the narrow "bottleneck" of two 4-gluon-coefficient combinations 
(\ref{z1}/\ref{z2}) appearing in only three of the 3-gluon conditions. This 
situation of a {\it near decoupling of the 4-gluon self-consistency problem} 
renders the following strategy sensible ( it is, in any case, the only
practical strategy at present ). One omits, as a first step, the 
4-gluon conditions completely, and treats the parameters $Z_1$, $Z_2$ 
appearing in (\ref{sce-3gl-3}-\ref{sce-3gl-5}) as two additional 
unknowns in the 2-point-plus-3-point-system, which thereby
becomes doubly underdetermined. Combined with the scaling property of
point (a), which effectively reduces the number of unknowns by one,
this results in an {\em effective one-parameter freedom} in the 
solution of the 2-plus-3-point problem. Since the number 
of calculable coefficients -- in the present case, fourteen in the 
2-and-3-point amplitudes, and seventeen in the "minimal" four-gluon vertex 
to be discussed in \cite{4} -- is much larger, the solutions will 
still be nontrivial and informative; in particular, one may explore
in what range, if any, of this one-parameter freedom there exist
physically acceptable solutions. 

In a second step, which we defer to \cite{4}, one may then 
adjoin the values of $Z_1$, $Z_2$ thus determined as additional 
constraints to the 4-gluon self-consistency problem: 
this will represent only a minor 
increase in the anyway massive overdetermination of that problem.
Since the 4-gluon system refers to the 2-and-3-point coefficients,
it inherits the effective one-parameter freedom, and to within
that freedom may be dealt with separately, with methods adapted
to its overdetermined nature.

It would seem that any one coefficient or combination of coefficients
of the 2-and-3-point system could be used to parametrize the
effective one-parameter freedom; in particular, some combination
of the quantities (\ref{z1}/\ref{z2}), which caused the freedom
in the first place, would seem to be a natural parameter. However,
one again faces unexpected restrictions here: due to the peculiar 
structure of the system in its fermionic unknowns, the fixing of a 
combination of non-propagator parameters, instead of rendering
the system well-determined, usually splits it into an over- and
an underdetermined part. The parametrizing quantity should thus refer
to propagator coefficients. In the following we will choose, 
for no other reasons than technical simplicity, the rescaled quark
self-energy coefficient $\tilde{w}_1$.

{\bf (d)}
At present, the question remains open as to whether there exists a
preferred or natural way of finally removing the one-parameter freedom.
One might think of recalculating the quantities $\tilde{Z}_1$,
$\tilde{Z}_2$ later from the least-squares four-gluon solution
and see if there is a parameter range where they agree, at least
qualitatively, with those from the 2-and-3-point solution. We shall
indeed do this in \cite{4}, but shall see that in the parameter
range where the entire solution is physically acceptable, a mismatch
is unavoidable at $r=1$, although small in the case of a pure gluon theory.
Alternatively, the vanishing of any of the previously noted 
approximation errors existing at the ($r=1$, $l=1$) level could 
be used as a condition. The common problem of all conditions of this 
kind is that (i) there are several of them, and any selection from among 
them appears arbitrary, (ii) they are mostly so restrictive that 
their imposition leaves only the trivial solution, with all 
nonperturbative coefficients vanishing.  The message 
the defect terms seem to convey is that the still rather simple and 
rigid structure of the $r=1$ system of approximants entails unavoidable 
approximation errors that cannot be forced to zero without overstraining 
that structure; they can disappear only gradually as $r$ is increased.
\subsection{Discussion of solutions}
The system augmented by (\ref{cond-ghost}) at a fixed value of $\tilde w_1$ 
may be reduced by successive elimination to an algebraic equation of the 
10th degree for the quantity $\tilde w_3$. The other coefficients can then 
be calculated recursively from the solutions of this equation and eq.
(\ref{cond-quark}), and depend parametrically on $\tilde w_1$. 
( These calculations have been performed using the MAPLE V 
computer-algebra system ). The following noteworthy features emerge.

{\bf (a)}
When assuming $x_1<0$ and performing the rescaling (\ref{tilde-coefs}/
\ref{tilde-lambda}) with $|x_1|=-x_1$ instead of $x_1$, the ten roots obtained 
for $\tilde{w}_3$ are all complex. Such solutions can immediately be discarded
as unphysical, since they lead to vertex functions not real at real Euclidean 
momenta, and the nonlinear nature of the system permits no superposition to
obtain real solutions. Therefore no physical solutions have been lost
by assuming $x_1>0$ and rescaling as in (\ref{tilde-coefs}/
\ref{tilde-lambda}).

{\bf (b)}
Over a range $0.3 \le \tilde{w}_1 \le 1.2$ ( all ranges quoted are 
approximate ) only eight of the ten $\tilde{w}_3$ roots 
come in complex-conjugate pairs, but two are real:
{\it there exist solutions with all vertex coefficients real}. This result
is entirely nontrivial, and represents substantial progress over the earlier 
attempt of refs. \cite{5}, where what we would now call the $r=1$ 
level of approximation was studied in a more heuristic fashion, with strong
a priori simplifications of the vertex approximants, and without taking
the compensating-poles mechanism into account. There, only partly
real solutions could be found. 

Of the two real $\tilde{w}_3$ roots, one is negative and one positive. The
negative $\tilde{w}_3$ value always turns out to lead to "tachyonic" pole
positions (negative values of the $\rho_{\pm}^2$ of (\ref{pro-pol}) below) 
in at least one of the two propagators, and can also be discarded as 
unphysical. Again it is nontrivial and noteworthy that only one 
of several solutions of the nonlinear system stands out 
as a candidate for a physical solution. Each of the two real solutions 
still exhibits the doubling of eqs. (\ref{oricoef}/\ref{mirrorcoef}), 
i.e. has a mirror solution for some of its fermionic parameters in the 
range $-1.2 \le \tilde{w}_1 \le -0.3$.
\begin{table}[!htb] 
\begin{center} 
\begin{tabular}[t]{|c||c|c|c|c|c|}\hline
             & & & & & \\ 
             & $0.3<\tilde{w}_1<0.4$ & $0.4<\tilde{w}_1<0.5$ 
                  & $0.5<\tilde{w}_1<0.9$ &
                      $0.9<\tilde{w}_1<1.1$ & $1.1<\tilde{w}_1<1.2$ \\ 
             & & & & & \\  \hline\hline
            \rule{0cm}{0.5cm} 
      $\tilde{\sigma}_+$ & $<\,0$ & \multicolumn{2}{|c|}{complex} & 
            $<\,0$ & $<\,0$ \\
            \rule{0cm}{0.5cm}
      $\tilde{\sigma}_-$ & $>\,0$ & \multicolumn{2}{|c|}{conjugate} & 
              $<\,0$ & $<\,0$ 
                                     \\ [0.2cm] \hline
            \rule{0cm}{0.5cm}
      $\tilde{\rho}_+^2$ & $>\,0$ & $>\,0$ & \multicolumn{2}{|c|}
                 {complex} & $>\,0$ \\
            \rule{0cm}{0.5cm}
      $\tilde{\rho}_-^2$ & $>\,0$ & $>\,0$ & \multicolumn{2}{|c|}
                 {conjugate} & $>\,0$ 
                                     \\ [0.2cm] \hline
\end{tabular}
\\[5mm] {\bf Table 5.1:} Ranges of interest for propagator-pole parameters
\end{center}
\end{table}

{\bf (c)}
For the solution in the range $0.3 \le \tilde{w}_1 \le 1.2$ with 
$\tilde{w}_3$ real and positive, table 5.1 records the nature of the poles
in the Euclidean transverse-gluon and fermion propagators, which now read
\begin{align}
      D_T^{[1,0]}(k^2) &=\frac{k^2+u_2\Lambda^2}{(k^2+\sigma_+ \Lambda^2)
                     (k^2+\sigma_- \Lambda^2)},         \label{glupro}   \\
      S^{[1,0]}(\slash p) &=\frac{\slash p+ w_2\Lambda}{(\slash p
              +\rho_+ \Lambda)(\slash p+\rho_- \Lambda)},
\end{align}
( see eqs. (\ref{gl-prop-sigma}) and (\ref{ferm-prop}) ).One finds that 
over a narrower range of $\tilde{w}_1$, namely, 
\begin{equation}
             0.5 \le \tilde{w}_1 \le 0.9                 \label{new-range}
\end{equation}
the solution with the real and positive $\tilde{w}_3$ root in addition 
fulfills the three inqualities
\begin{equation}
\gamma_V^2\equiv u_3-\left(\frac{u_1-u_2}{2}\right)^2>0,\qquad
\gamma_F^2\equiv w_3-\left(\frac{w_1-w_2}{2}\right)^2>0,\qquad
\left|w_1+w_2\right| > 0, \label{conf-cond}
\end{equation}
so that {\it both propagators simultaneously} exhibit complex-conjugate 
pole pairs at $k^2=-\sigma_{\pm}\Lambda^2$ and $p^2=-\rho^2_{\pm}\Lambda^2$,
where
\begin{equation} \label{pro-pol}
        \sigma_{\pm}=\frac{u_1+u_2}{2} \pm i\gamma_V,\qquad
        \rho^2_{\pm}=\frac{w_1^2+w_2^2}{2} -w_3 \pm i (w_1+w_2)\gamma_F.
\end{equation}
Thus there exist solutions in which the elementary excitations of the two 
basic QCD fields are both short-lived. In the present framework this is 
the essential indicator of confinement, since it implies the vanishing
of S-matrix elements with external single-gluon or single-quark legs
\cite{5}. We again regard it as nontrivial that a parameter range should at
all exist in which this situation prevails. Note also that in the gluonic
portion of table 5.1, there is always at least one ``tachyonic'' 
gluon-propagator pole outside the slightly wider range $0.4 \leq 
\tilde{w}_1 \leq 0.9$ ); we view it as significant that the only solutions
with real vertices and non-tachyonic gluons have gluon propagators with
complex pole pairs. -- Over the range (\ref{new-range}), 
all other vertex coefficients are only weakly varying. 

This interesting solution still has a "mirror" solution in the sense of
(\ref{mirrorcoef}), i. e. in a $\tilde{w}_1$ range which is the negative of 
(\ref{new-range}). We are not aware of a {\it theoretical} criterion that
would resolve this discrete ambiguity. One might prefer the solution 
with negative $\tilde{w}_1$ on the {\it empirical} grounds that it
gives the $r=1$, zeroth-order fermion condensate
\begin{equation}
    \Big( \, g_0^2 \,\big< 0|\bar{\Psi}\Psi |0 \big> \,\Big)^{[1,0]}
           = \  \frac{12}{\beta_0} \ \big( w_1^{\,3} - 2w_1w_3 - 
                      2w_2w_3 \big) \ \Lambda^3      \label{fermcond}
\end{equation} 
the negative sign established in the context of current algebra und QCD sum 
rules. However, there is no reason for believing that a solution with
low $r$, which is generally crude and more so in the fermion sector, 
must already give the correct sign for such a sensitive quantity.

{\bf (d)}
Upon imposing condition (\ref{cond-gluon}), from the gluonic equation of 
motion, to remove the residual freedom in the $\tilde{w}_1$, we find values 
$\tilde{w}_1$ not only outside the range (\ref{new-range}) but in fact 
outside the larger range of table 5.1\ , where $\tilde{w}_3$ and thus the 
entire solution creases to be real and physically acceptable: enforcing 
(\ref{cond-gluon}) one loses the possibility of a physical solution and of 
all the features noted in (c). Thus (\ref{cond-gluon}), like removal of the 
order-$g^2$ defect terms noted above, seems to be a strongly restrictive 
condition that the simple $r=1$ structure is too rigid to accommodate.  

Within the general strategy suggested and used here, it appears that
restriction to the quite limited parameter range where all propagator
singularites are complex conjugate, and none tachyonic, in itself
represents a sensible limitation to the one-parameter freedom, and
one that is difficult to narrow further without overburdening the
$r=1$ approximation.
\begin{table}[!htb] 
\vspace{4mm}
\begin{center}
\begin{tabular}{|ccc|ccccc|c|}
\hline
$\tilde u_1$ & $\tilde u_2$ & $\tilde u_3$ &
$\tilde x_1$ & $\tilde x_2$ & $\tilde x_3$ & $\tilde x_4$ & $\tilde x_5$ &
$\gamma_V$ \\
\hline\hline
$-0.3604$ & $-0.4884$ & $+0.1299$ & $+1.0000$ & $-8.7433$  
                         & $+8.9088$ & $-3.2607$ & $-6.2711$ & $0.3547$   \\
\hline
\end{tabular}
\vspace{4mm}
\begin{tabular}{|ccc|ccccc|c|}
\hline
$\tilde w_1$ & $\tilde w_2$ & $\tilde w_3$ &
$\tilde z_1$ & $\tilde z_2$ & $\tilde z_3$ & $\tilde z_4$ & $\tilde z_5$ &
$\gamma_F$ \\
\hline\hline
$+0.6749$ & $+0.6749$ & $+0.1202$ & $-0.9561$ & $-0.9356$  
                   & $-0.4282$ & $+0.4094$ & $+0.2242$ & $0.3468$        \\
\hline
\end{tabular}
\\[3mm]        {\bf Table 5.2:} Typical solution with $N_F=2$
\end{center}
\end{table}

{\bf (e)}
For use of the generalized Feynman rules in applications, we list 
in table 5.2 a typical set of two-point and three-point vertex 
coefficients for $N_F=2$, for $\tilde w_1$ chosen in about the middle 
of the range (\ref{new-range}). ( Since in that range
$\tilde w_2$ varies slowly and is itself of modulus $\approx$ $0.7$, 
we choose the point $\tilde w_1=\tilde w_2$ for simplicity.) This set still 
needs completion through the corresponding four-gluon vertex coefficients 
$\zeta_i$, but since many lower-order calculations need at most 3-point 
vertices, it seems legitimate to defer presentation of these to \cite{4}.
The propagator-pole parameters (\ref{pro-pol}) for this solution are
\begin{equation}
      \sigma_{\pm} = (-0.4245 \pm i 0.3547) \tilde{\Lambda}^2 \; , \qquad
      \rho_{\pm}^2 = (0.3353 \pm i 0.4679) \tilde{\Lambda}^2 \; .
\end{equation}

For purposes of comparison, we also briefly look at solutions for the
pure-gluon theory ($N_F=0$). Here the parametrizing quantity may be 
taken to be the gluonic vertex coefficient $\tilde{x}_3$, and again
we choose a typical value, $\tilde{x}_3 \approx 1$, from the ( again 
existing ) range in which the gluonic propagator poles are 
complex conjugate. ( The value $\tilde x_3 = 1$, incidentally, is also one
which symmetrizes, though not removes, the defects of (\ref{3g-pert-defect})
in the perturbative three-gluon divergence ). Table 5.3 lists coefficients 
for this case.
\begin{table}[!hbt] 
\vspace{4mm}
\begin{center}
\begin{tabular}{|ccc|ccccc|c|}
\hline
$\tilde u_1$ & $\tilde u_2$ & $\tilde u_3$ &
$\tilde x_1$ & $\tilde x_2$ & $\tilde x_3$ & $\tilde x_4$ & $\tilde x_5$ &
$\gamma_V$ \\
\hline\hline
$-1.7429$ & $+0.8456$ & $+3.0376$ & $+1.0000$ & $-6.1825$  
                         & $+1.0000$ & $-4.8682$ & $+28.605$ & $1.1650$  \\
\hline
\end{tabular}
\\[4mm]       {\bf Table 5.3:} Typical solution for pure-gluon system
\end{center}
\end{table}
It is still impossible here to accommodate condition (\ref{cond-gluon}) 
in a physically acceptable solution. Note however that this solution 
is now unaffected by the doubling of (\ref{mirrorcoef}). 

The presence or absence of the massless-quark loops obviously has 
a strong effect on several of the coefficients. In particular, the
value of the transverse-gluon propagator function (\ref{glupro}) 
at $k^2=0$ --  a finite constant whose sign is determined by the $u_2$ 
parameter --  is positive for the pure-gluon system but negative in
the presence of the light quarks. It is again unlikely that the crude
$r=1$ solution should describe this effect quantitatively, but its 
qualitative trend is plausible from the minus signs of the fermionic
self-energy loops.

\section{Conclusion}
\setcounter{equation}{0}

We have demonstrated the feasibility of a self- consistent determination
of generalized Feynman rules, accounting for the nonperturbative
$\Lambda$ dependence of correlation functions through a modified
iterative solution, at the simplest level of systematic approximation
of that dependence. We have shown, and regard it as nontrivial,
that the nonlinear self-consistency problem admits physically
acceptable solutions, that these stand out clearly against a majority
of unphysical ones, and that there exist solutions in which both of
the elementary excitations of the basic QCD fields exhibit the short-ranged
propagation described by complex-conjugate propagator poles. 

It is useful to recall the restrictions under which we have studied this
self-consistency problem. We have considered
the $r=1$ level of rational approximation of the $\Lambda$ dependence
-- the lowest level of interest for a "confining" theory like QCD. 
The limitations inherent in this low approximation order have 
become clearly visible; its structure is far too simple and rigid
to satisfy all desirable conditions and restrictions simultaneously.

We have worked with the "ordinary" DS equations only, with bare 
vertices on their distinguished, left-hand external legs. We have evaluated 
the self-consistency conditions on the one-loop level, in Landau gauge,
and with a special decoupling of the 4-gluon-conditions as suggested
by the peculiar "bottleneck" structure of the system. It is desirable for
future work to gradually remove these limitations, and in particular 
to study the Bethe-Salpeter-resummed forms of the vertex equations, which
may have more of the important physical effects shifted into the low loop 
orders.

Even with such improvements, two problems are certain to persist that have
emerged clearly from the present study. One, which arises only when
studying vertices with at least three legs, and has nothing to do with the 
specifics of the present method, is the overdetermination dilemma unavoidable 
when seeking approximate-but-symmetric solutions to the not manifestly 
symmetric DS equations. This may be ``swept under the rug'' by trivial 
symmetrizations, but only at the expense of depriving oneself of an important 
measure of error. The second problem is that the DS equations, through
their divergent parts, do not fix the common set of denominator parameters
of the approximants. A way of understanding this interesting result is
to recall the relation with the operator-product expansion, as discussed
in sect. (2.3) of \cite{3}: the OPE, in its higher orders, contains
vacuum expectations of local operators of arbitrarily high compositeness,
whereas the DS equations contain at most insertions of three operators at the 
same spacetime point. The extra composite-operator renormalizations
required by those higher condensates represent extra information which the
usual DS system does not supply directly. Equation-of-motion condensate 
conditions,
which do represent statements about quantities of higher compositeness,
are capable of supplying the extra information, and natural complements
insofar as they are special, zero-separation cases of DS equations.
Their role, which in the present context may still have looked marginal,
will clearly become more central when going to higher levels $r$.
 
It should be kept in mind that for a confining field system such as QCD,
the generalized Feynman rules as considered here allow only the calculation
of off-shell Green's functions of the elementary fields. These still carry 
little observable information, although the spectrum of the elementary
excitations as determined by the singularities of the two-point-functions
does constitute important {\it qualitative} information. To calculate
on-shell amplitudes, whose external legs are bound states, one would need
in addition bound-state vertices to sit at the outer corners of S-matrix
diagrams. These have not been touched upon in this paper, since they are
conceptually quite different from the zeroth-perturbative-order quantities:
they arise from partial (ladder or improved-ladder) resummation of
quasi-perturbative corrections $g^{2p}\,\Gamma_N^{(r,p)}$, with $p\ge 1$,
for certain superficially {\it convergent} amplitudes $\Gamma_N$, in which
the mechanism of eq.\ (\ref{pi-eps}) plays no role. Their determination
must therefore rely on the established Bethe-Salpeter methods for bound
states.

On the other hand we do believe that the calculations described here achieve
something new by dealing with the complete set of superficially divergent
QCD vertices in one consistent approximation, and that they demonstrate
a nontrivial, renormalization-related way of how the 
renormalization-group invariant mass scale establishes itself in the
correlation functions of an asymptotically free theory. 
\vspace{6cm}
\begin{center}    {\bf Acknowledgment}    \end{center}
\noindent
One of the authors (M.S.) is grateful to D. Sch\"utte for an invitation
and for support to attend the 1997 Bonn Workshop on Confinement where
parts of this paper were written.

\newpage
\begin{center}

\end{center}
%
%
%

\vspace*{-8cm}

\begin{figure}[t] 
\begin{center}
\epsfxsize=20cm \epsfbox{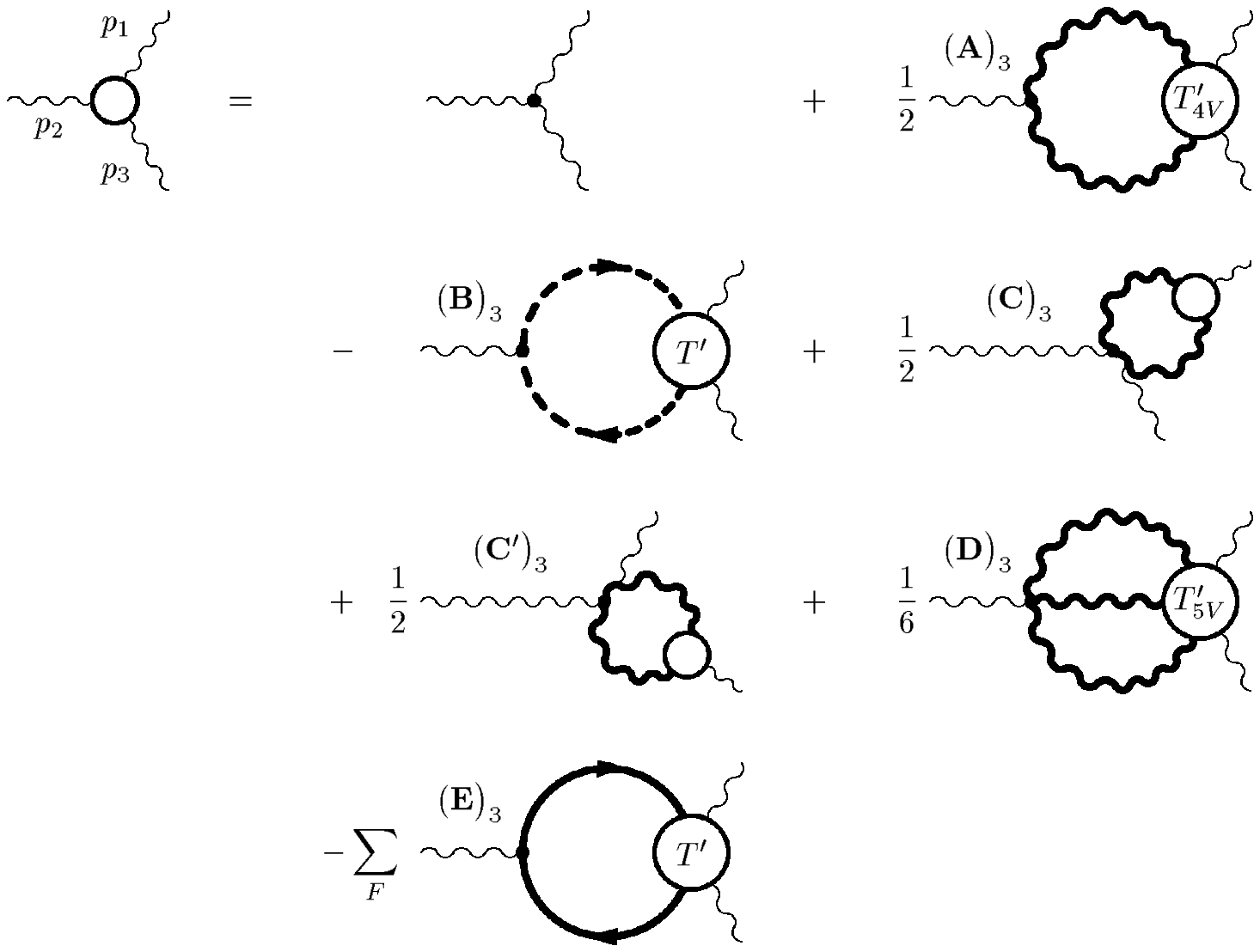}
\vspace{-8cm}

\caption{DS equation for the $\Gamma_{3V}$ Vertex.}
\end{center}
\end{figure}
\begin{figure}[t] 
\begin{center}
\epsfxsize=11cm \epsfbox{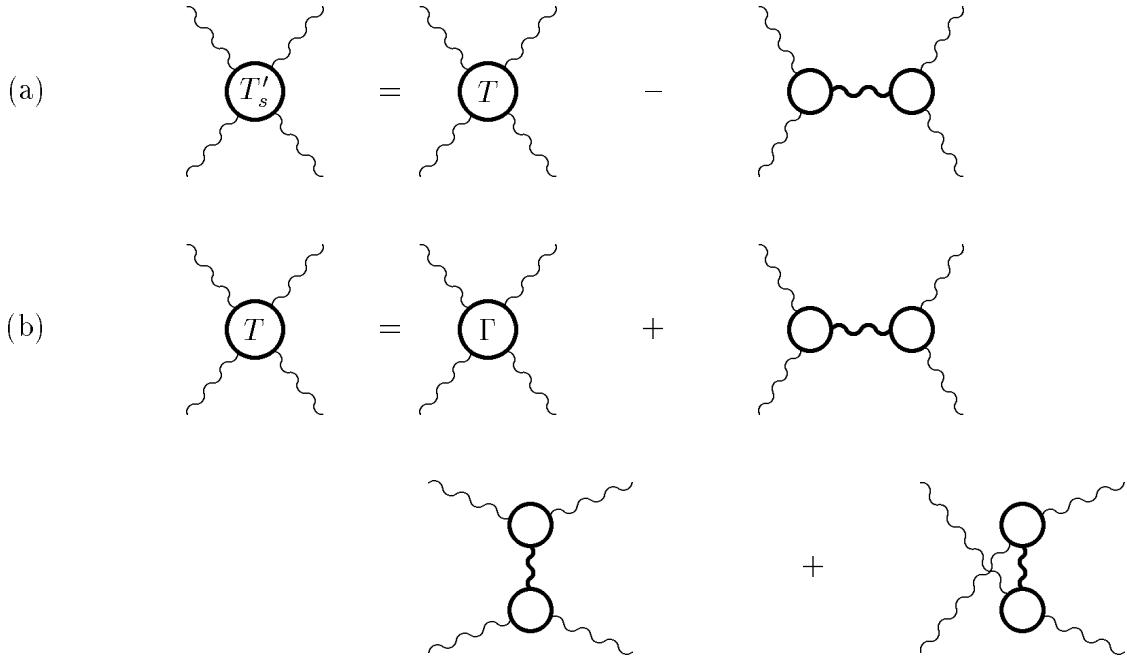}
\vspace{7mm}
\caption{Decomposition of the four-gluon $T$-matrix.}
\end{center}
\end{figure}

\begin{figure}[!hb] 
\vspace{1cm}
\begin{center}
\epsfxsize=14cm \epsfbox{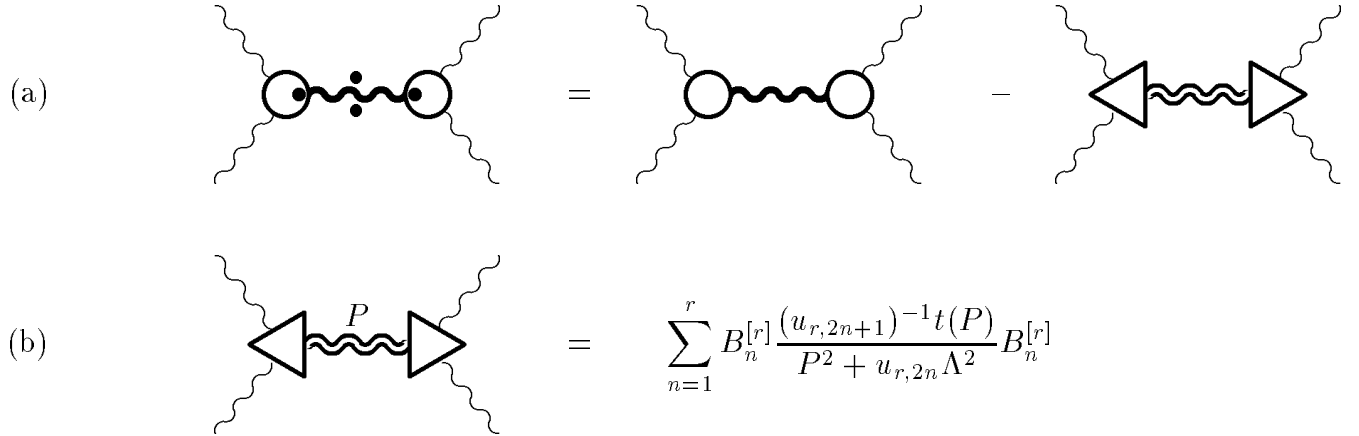}
\vspace{7mm}
\caption{"Softened" exchange diagram (a) and diagrammatic representation 
of the compensating pole (b).}
\end{center}
\end{figure}

\vspace*{-7cm}

\begin{figure}[t] 
\begin{center}
\epsfxsize=16cm \epsfbox{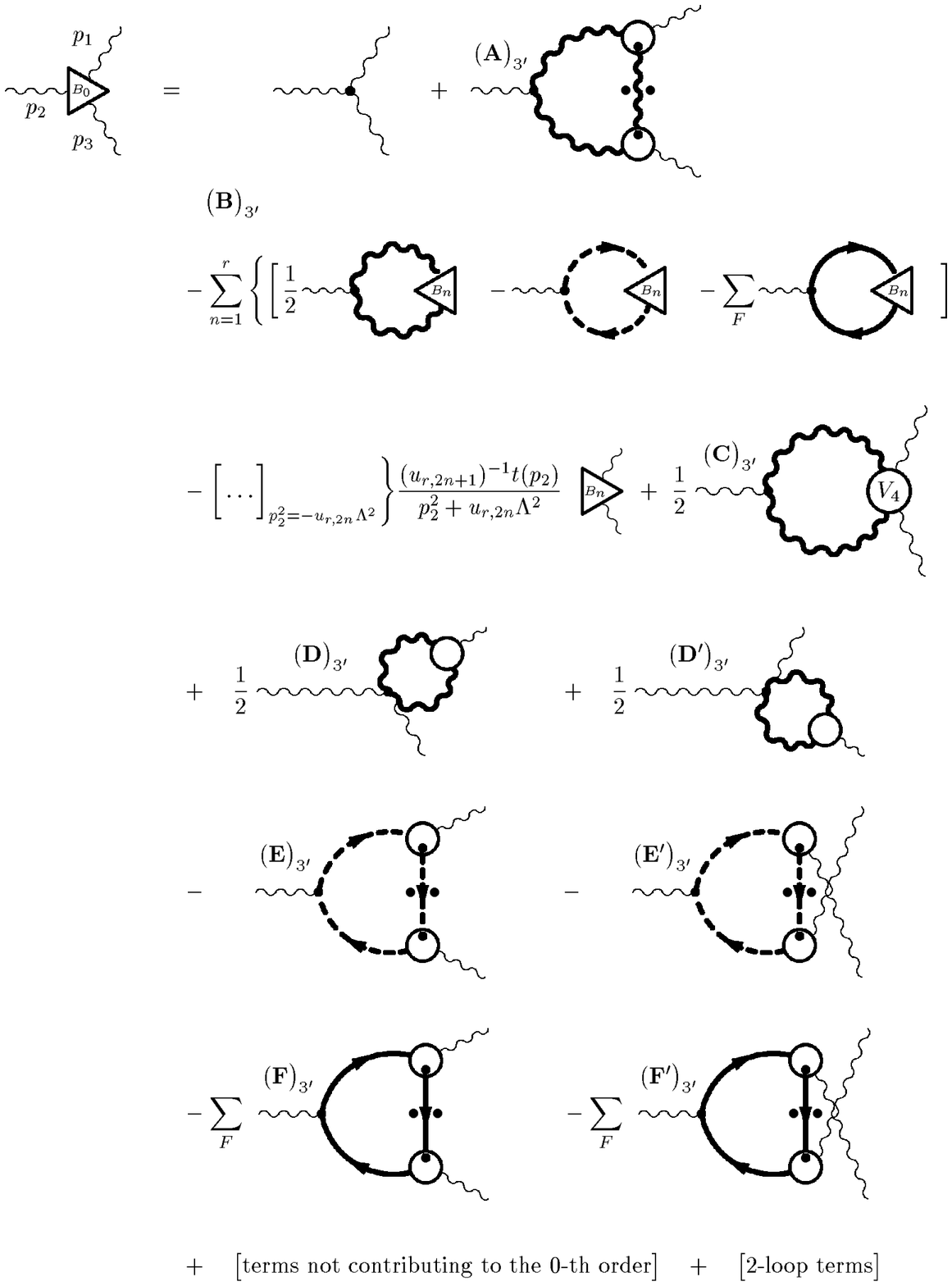}
\vspace{-1cm}

\caption{DS equation for the $B_0$ part of the $\Gamma_{3V}$-Vertex.}
\end{center}
\end{figure}
\begin{figure}[t] 
\begin{center}
\epsfxsize=16cm \epsfbox{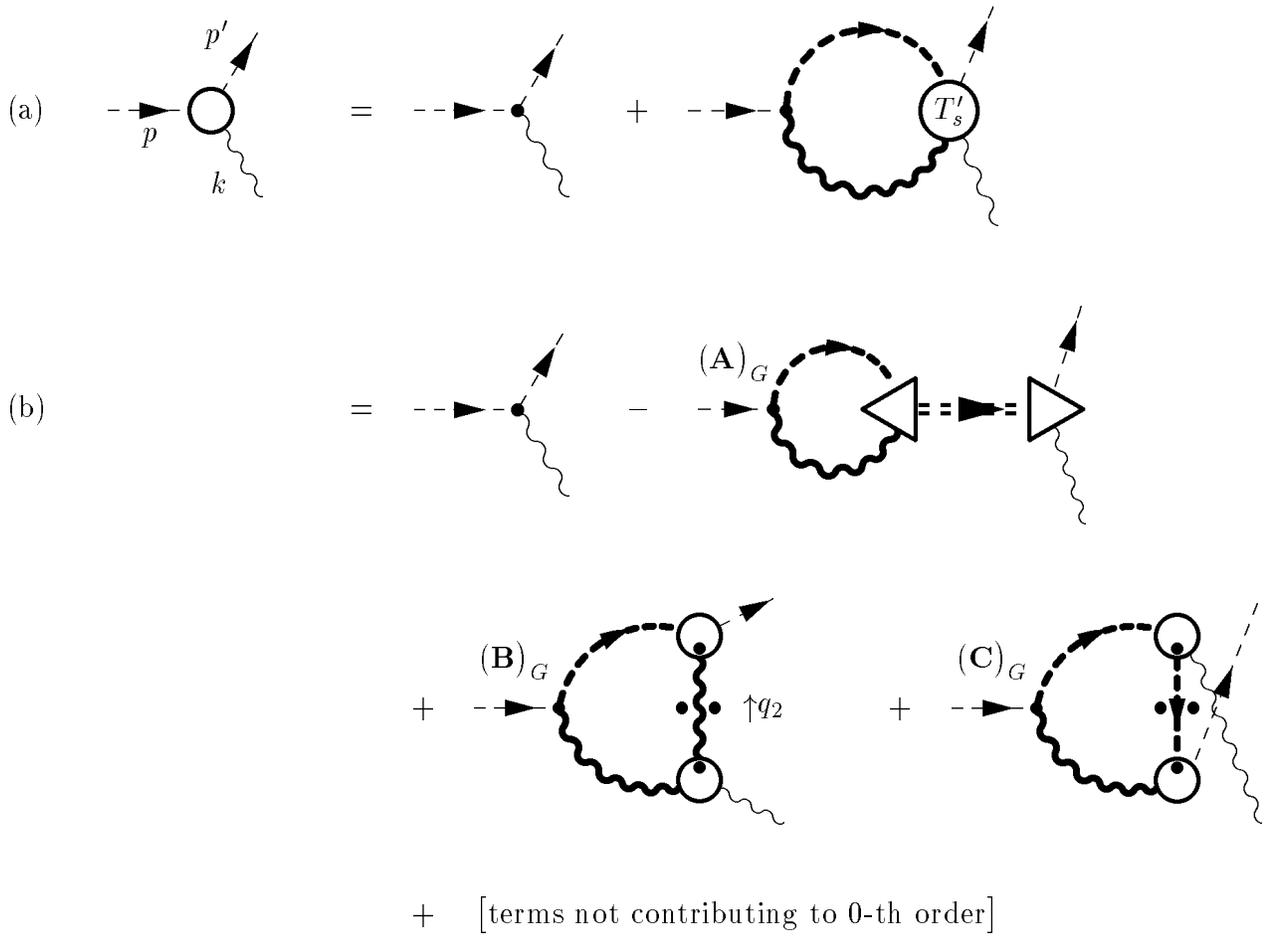}
\vspace{5mm}
\caption{DS equation for the ghost-gluon 3-point vertex in the ghost channel.}
\end{center}
\end{figure}

\vspace{1.5cm}
\begin{figure}[!hb] 
\hspace{2cm}
\epsfxsize=8cm \epsfbox{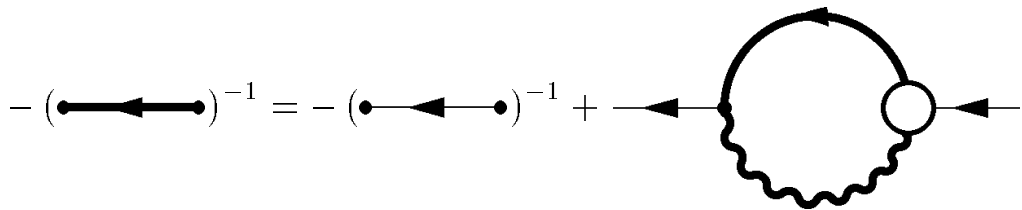}
\vspace{5mm}
\caption{DS equation for the inverse fermion propagator.}
\end{figure}

\vspace*{-3cm}

\begin{figure}[t]
\begin{center}
\epsfxsize=20cm \epsfbox{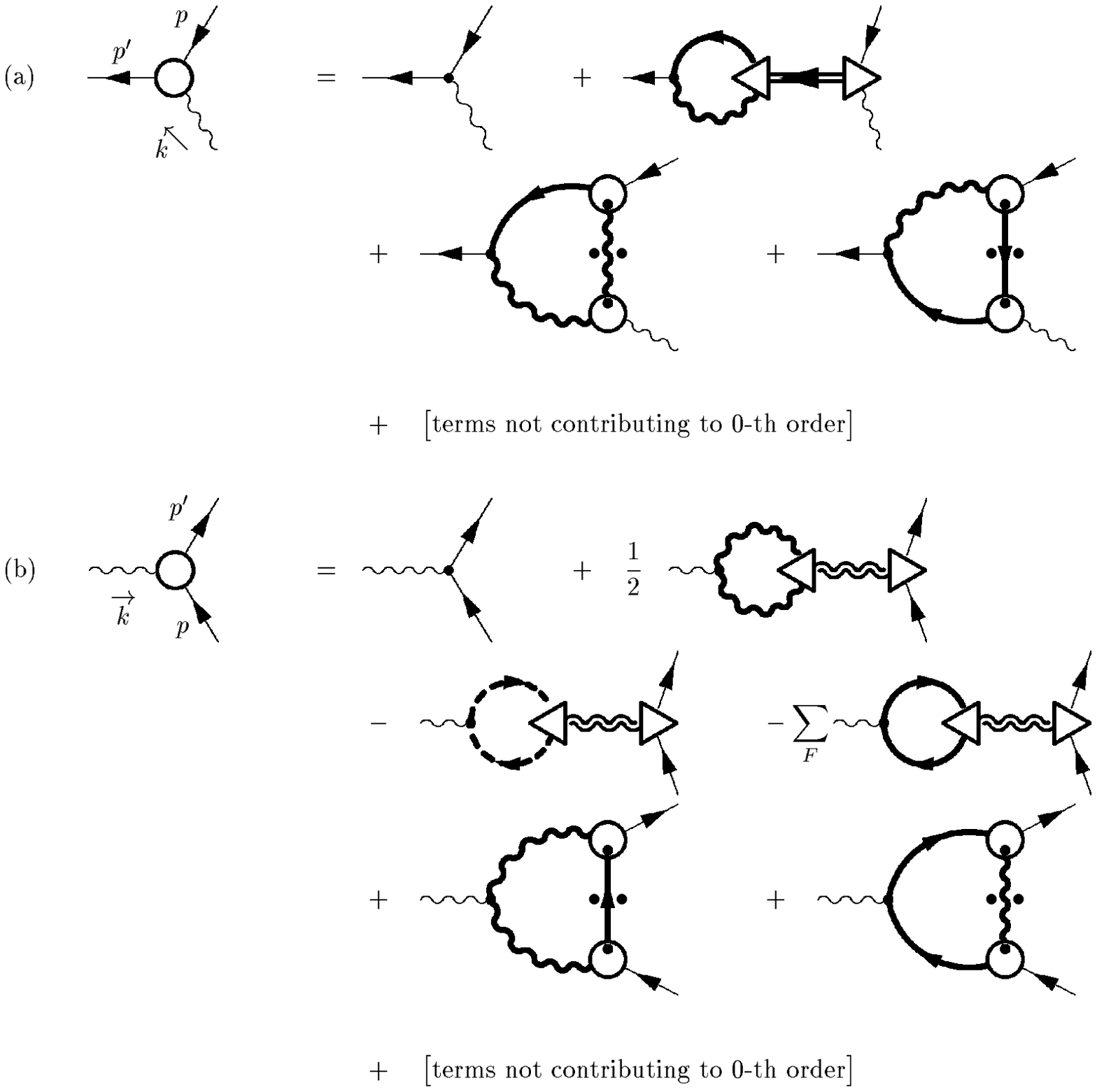}

\vspace{-7cm}

\caption{Equivalent DS equations for the quark-antiquark-gluon vertex 
in (a) fermionic channel and (b) gluonic channel.}
\end{center}
\end{figure}


\begin{thebibliography}{99}

\bibitem[1]{1} D.\ J.\ Gross and A.\ Neveu, Phys.\ Rev.\ {\textbf{D10}} (1974)
 3235; C.\ G.\ Callan, R.\ F.\ Dashen, and D.\ J.\ Gross, Phys.\ Rev.\ 
{\textbf{D17}} (1978) 2717  
\bibitem[2]{2} G.\ t'Hooft, in: A.\ Zichichi (Ed.), {\it The Whys of
          Subnuclear Physics} (Proceedings Erice 1977), Plenum, New York, 1979
\bibitem[3]{3} M.\ Stingl, Z. Physik {\textbf{A353}} (1996) 423, 
               E-print archive: hep-th/9502157 
\bibitem[4]{4} L.\ Driesen and M.\ Stingl, Extended Iterative Scheme for QCD:
         the Four-Gluon Vertex, preprint MS-TPI-98-19 (August 1998),
         to be submitted to E-print archive hep-th 
\bibitem[5]{5} U.\ H\"abel, R.\ K\"onning, H.-G.\ Reusch, M.\ Stingl, 
          S.\ Wigard, Z.\ Physik {\textbf{A336}}, (1990) 423, 435
\bibitem[6]{6} C.\ D.\ Roberts and A.\ G.\ Williams, Prog.\ Part.\ and 
          Nucl.\ Phys.\ {\textbf{33}} (1994) 477
\bibitem[7]{7} R.\ Jackiw and K.\ Johnson, Phys.\ Rev.\ {\textbf{D8}}, (1973)
               2386, J.\ M.\ Cornwall and R.\ E.\ Norton, Phys.\ Rev.\ 
              {\textbf{D8}}, (1973) 3338
\bibitem[8]{8} W.\ Zimmermann, Nuovo Cim.\ {\textbf{13}} (1959) 503, 
           {\textbf{16}} (1960) 690 
\bibitem[9]{9} J.\ Fromm, Diploma thesis, University of M\"unster, 1996 
               (unpublished); L.\ Driesen, Dr.\ rer.\ nat.\ thesis, University
               of M\"unster, 1997 (unpublished), J.\ Kuhrs, Dr.\ rer.\ nat.\ 
               thesis, University of M\"unster, 1997 (unpublished)   

\end{thebibliography}
\end{document}